\documentclass[fleqn,usenatbib, onecolumn]{mnras}

\usepackage{newtxtext,newtxmath}
\usepackage{xcolor}

\usepackage[left]{lineno}
\usepackage{epstopdf}
\usepackage{ulem}
\usepackage{times}
\usepackage{graphicx}
\usepackage[usenames,dvipsnames]{pstricks}
\usepackage{psfrag}
\usepackage{lipsum}
\usepackage{natbib}

\usepackage[T1]{fontenc}

\DeclareRobustCommand{\VAN}[3]{#2}
\let\VANthebibliography\thebibliography
\def\thebibliography{\DeclareRobustCommand{\VAN}[3]{##3}\VANthebibliography}

\usepackage{graphicx}	
\usepackage{amsmath}	

\def\araa{ARA\&A}
\def\apj{ApJ}
\def\apjl{ApJ}
\def\apjs{ApJS}
\def\ao{Appl.~Opt.}
\def\aap{A\&A}
\def\mnras{MNRAS}
\def\pasj{PASJ}
\def\nat{Nature}
\def\physrep{Phys.~Rep.}

\newcommand{\be}{\begin{equation}}
\newcommand{\ee}{\end{equation}}
\newcommand{\bary}{\begin{eqnarray}}
\newcommand{\eary}{\end{eqnarray}}

\title[Emergence of magnetic fields]{Magnetic Burial in Millisecond Magnetars and Late GRB Afterglow Signatures}

\author[Fraija, N. et al.]{
Nissim Fraija,$^{1}$\thanks{E-mail: nifraija@astro.unam.mx}
C. G. Bernal,$^{2}$
A.~ Galv\'{a}n$^{3}$
B.~ Betancourt Kamenetskaia$^{4,5}$
M.G. Dainotti$^{6,7}$
\\
$^{1}$Instituto de Astronom\'ia, Universidad Nacional Aut\'{o}noma de M\'{e}xico, Apdo. Postal 70-264, Cd. Universitaria, Ciudad de M\'{e}xico 04510\\
$^{2}$Universidad Nacional de Colombia, Apdo. Postal 111321, Cd. Universitaria, Bogotá\\
$^{3}$Instituto de F\'isica, Universidad Nacional Aut\'{o}noma de M\'{e}xico, Apdo. Postal 70-264, Cd. Universitaria, Ciudad de M\'{e}xico 04510\\
$^{4}$Technical University of Munich, TUM School of Natural Sciences, Physics Department, James-Franck-Str 1, 85748 Garching, Germany\\
$^{5}$Max-Planck-Institut f\"ur Physik (Werner-Heisenberg-Institut), Boltzmannstr. 8, 85748 Garching, Germany\\
$^{6}$Division of Science, National Astronomical Observatory of Japan, 2-21-1 Osawa, Mitaka, Tokyo 181-8588, Japan\\
$^{7}$The Graduate University for Advanced Studies (SOKENDAI), 2-21-1 Osawa, Mitaka, Tokyo 181-8588, Japan\\
}

\date{Accepted XXX. Received YYY; in original form ZZZ}

\pubyear{\the\year{}}

\begin{document}
\label{firstpage}
\pagerange{\pageref{firstpage}--\pageref{lastpage}}
\maketitle

\begin{abstract}
 Millisecond magnetars, one of the potential candidates for the central engine of Gamma-ray bursts (GRBs), can experience significant magnetic field enhancement shortly after their formation. In some cases, this evolution is further influenced by the accretion of stellar debris, which modifies the dipole magnetic field strength. During a hypercritical accretion phase that lasts seconds or longer after the progenitor explosion, a thin crust may form, submerging the magnetic field (the so-called magnetic burial scenario). Once hypercritical accretion ceases, the buried field can diffuse back through the crust, delaying the external dipole's reactivation. On the other hand, observations have shown that relativistic outflows ejected by these objects and decelerated by the circumburst environment cause a late and temporary emission known as afterglow. This work investigates how the submergence and subsequent reemergence of the magnetar magnetic field, on a few years timescales, affect the GRB afterglow dynamics. Specifically, we apply this phenomenological scenario to the late-time X-ray excess observed approximately three years post-burst in GW170817/GRB 170817A, exploring how the evolving magnetic field strength may contribute to this emission.
 Our modelling of GRB\,170817A indicates that $\gtrsim90$ percent of the external dipole flux was initially buried, re-emerging on a timescale $\tau_{B}=3-40$ yr and restoring a surface field $B\simeq(2-5)\times10^{15}\,$G; the late–time X-ray brightening is far better reproduced by this scenario than by models without burial.
\end{abstract}





\section{Introduction}

Gamma-ray bursts (GRBs), one of the most captivating astrophysical transients, are the brightest explosions in the Universe. Gamma-ray observations have shown that the merger of two NSs (NSs) can produce short GRBs, lasting less than a few seconds \citep{1992ApJ...392L...9D, 1992Natur.357..472U, 1994MNRAS.270..480T, 2011MNRAS.413.2031M}, while core collapses (CCs) of massive, dying stars can produce long GRBs that persist for over a few seconds \citep{1993ApJ...405..273W,1998ApJ...494L..45P,Macfadyen+99col,Woosley2006ARA&A}. However, the nature of the central compact object that powers the GRBs remains uncertain. An accreting black hole \citep[BH,][]{1992ApJ...395L..83N, 1993ApJ...405..273W, 1999ApJ...526..152F, 1998ApJ...494L..45P, 1999ApJ...524..262M, 2008MNRAS.385L..28B} or an extremely fast rotating,  highly magnetized NS \citep[a millisecond magnetar;][]{1992ApJ...392L...9D, 1992Natur.357..472U, 1994MNRAS.270..480T, 2011MNRAS.413.2031M} represent the two leading models for the central GRB engines discussed in the literature.  In the first scenario, the CC of a massive star or the merger of two NS leaves material as a torus. This torus can persist for a long time \citep[e.g., see][]{1993ApJ...405..273W}. Relativistic outflows are powered by the energy generated during accretion or accretion-mediated extraction from a Kerr BH \citep{1977MNRAS.179..433B}. In the magnetar scenario, a millisecond NS forms with sufficient rotational energy to temporarily prevent collapse \cite[e.g., see][]{2011MNRAS.413.2031M}.

The magnetic dipole spin-down brightness of the single-object magnetar drives the relativistic outflow. Whether the compact object is a BH or a magnetar, both CC-SNe and NS mergers are expected to leave behind progenitor remnants. A fraction of this material forms an accretion disk and gradually falls back onto the compact object \citep{1989ApJ...346..847C, 2012ApJ...752...32W, 2012MNRAS.419L...1Q}.  In the magnetar scenario, accretion predominantly contributes more angular momentum to the system, thus growing the reservoir of rotational energy available to propel the relativistic outflow \citep{2011ApJ...736..108P}. When the mass of the magnetar surpasses a specific threshold, differential rotation can no longer maintain its stability, resulting in the collapse of the magnetar into a BH \citep{2010MNRAS.409..531R}.  Independent of the progenitor type, observations show that highly relativistic collimated outflows produce the afterglow emission when they come into contact with the external environment, accelerating nonthermal electrons, which cool mainly by synchrotron radiation \citep{2002ApJ...571..779P, 2004ApJ...609L...1T, 2015PhR...561....1K}.

In addition, recent work by \citet{SuvorovKokkotas2020} argues that the shallow X-ray plateau can also arise from free precession of the newborn
magnetar, driven by a stable toroidal field that tilts the symmetry
axis away from the spin axis.  In this picture the plateau is only
quasi-flat: small-amplitude modulations on the precession
period (\(\sim10^{3}\)–\(10^{4}\)~s) are expected, together with a
long-lived, nearly monochromatic gravitational-wave signal. The present work focuses instead on magnetic submergence and
reemergence; we return to a comparison of the two mechanisms in the Conclusion Section.
\\

Millisecond magnetars likely undergo significant magnetic field amplification following their formation, influencing the dynamics and emissions observed in some GRBs. Several observational and theoretical evidence supports this scenario, including early dynamo mechanisms \citep{1992ApJ...392L...9D, 1993ApJ...408..194T}, magnetohydrodynamic (MHD) simulations \citep{2006Sci...312..719P, 2013ApJ...771L..26G, Bernal2013}, and the so-called "plateau" phase in GRB afterglows \citep{2013MNRAS.430.1061R, 2014ApJ...785...74L}. Observationally, the canonical X-ray light curve of a GRB typically exhibits an initial abrupt decay, followed by a shallow (plateau) segment, a normal decay, and eventually a late steep decay \citep[e.g., see][]{2006ApJ...642..354Z, 2006ApJ...642..389N}.  The plateau phase is often interpreted as continuous energy injection from the central engine, either through fallback accretion onto a BH \citep{Kumar2008,Cannizzo2009,cannizzo2011,2018ApJ...857...95M} or through the spin-down luminosity of a newborn magnetar \citep{zhang2001,2007ApJ...665..599T,2018ApJ...857...95M}. Once hypercritical accretion ceases,\footnote{Hereafter, we adopt the term "submergence of the magnetic field"  \citep[e.g., see,][]{art_dany, geppert1999submergence, 2013ApJ...770..106B} instead of "magnetic burial," as used by some authors \citep[e.g., see][]{2002MNRAS.332..933C, 2017JApA...38...48M}. Notably, other authors have also used the term "the hidden magnetic field scenario" to describe the same phenomenon \citep[e.g., see][]{2012MNRAS.425.2487V, 2016MNRAS.456.3813T}.} the buried (or "submerged") magnetic field can begin to diffuse back through the crust, leading to a delayed reactivation of the external dipole. This process, originally proposed for pulsars by \cite{michel1994magnetic} and expanded by \cite{art_dany}, depends primarily on the amount of accreted material. As shown by \cite{geppert1999submergence, konar2017magnetic, mukherjee2017revisiting}, sufficient mass accumulation can keep a young NS effectively unmagnetized at the surface for decades to centuries. Field reemergence may then be driven by mechanisms such as magnetic reconnection, turbulent dynamos, or thermomagnetic instabilities: processes that are complex and sensitive to the thermal and compositional profiles of the NS, often requiring detailed numerical simulations to be characterized \citep{art_dany, geppert1999submergence, vigelius2009resistive}.

In this work, we develop an analytical framework to estimate how buried field reemergence, on timescales of a few years, affects the spin-down of a newborn millisecond magnetar and, consequently, the late afterglow of GRBs. Specifically, we investigate how hypercritical accretion leads to field submergence, the subsequent reemergence of the dipole component, and its impact on the long-term evolution of the afterglow. As a concrete application, we examine the X-ray excess observed approximately three years post-burst in GW170817/GRB 170817A and explore whether an intensifying magnetar field could account for this late-time feature.
The subsequent sections presented in this work are structured as follows: Section \ref{sec2:model} introduces the model for field growth. Section \ref{sec3:model} derives the synchrotron light curves under reemergence conditions. Section \ref{sec4:model} applies our analytical framework to the afterglow observations of GRB 170817A. Finally, Section \ref{sec5:model} summarizes our main results and provides concluding remarks.


\section{Analytical Study of the Magnetic Field Growth Effect}\label{sec2:model}

\subsection{On the Magnetic Field Submergence}

As discussed above, millisecond magnetars likely undergo substantial magnetic field growth shortly after their formation, affecting both the dynamics and emission properties of some GRBs. This scenario is supported by multiple lines of observational and theoretical evidence: (i) Dynamo mechanisms -- Shortly after NS forms, either from a core-collapse supernova or a binary NS merger, differential rotation and convection can rapidly amplify the magnetic field to magnetar-level strengths. Millisecond spin periods are essential to enable these dynamo processes \citep{1992ApJ...392L...9D, 1993ApJ...408..194T}; (ii) MHD simulations -- Numerical studies show that during a NS merger or hyperaccretion phase, intense shear and compression can twist and amplify magnetic field lines. This rapid amplification contributes to the development of magnetar-level fields in the remnant \citep{2006Sci...312..719P, 2013ApJ...771L..26G, Bernal2013}; (iii) GRB afterglows and plateaus -- Many GRBs exhibit light curves characterized by an initial steep decay, followed by a shallow plateau, a normal decay, and finally a late steep decay. The plateau phase is commonly interpreted as the result of sustained energy injection from a rapidly spinning millisecond magnetar. In this scenario, the decaying magnetic field and spin-down of the magnetar supply energy over an extended period, shaping the afterglow light curve \citep{2013MNRAS.430.1061R, 2014ApJ...785...74L}; (iv) Energetics and spectral evolution -- The exceptionally high isotropic energies observed in some GRBs may be explained by the efficient conversion of a millisecond magnetar rotational kinetic energy into electromagnetic radiation \citep{2011MNRAS.413.2031M}. Similarly, changes in GRB spectral features can reflect an evolving central engine, where magnetic field reconfiguration and spin-down modify the underlying radiative processes \citep{zhang2001}; (v) NS mass, equation of state, and magnetar-quakes -- The ability of a newly formed NS to sustain rapid rotation and a strong magnetic field without collapsing into a black hole depends on its mass and equation of state parameters constrained by gravitational wave observations and pulsar timing \citep{2016ARA&A..54..401O}. Observational evidence of sudden flares from Galactic magnetars suggests that magnetic energy reconfiguration or partial release (the so-called magnetar-quakes) can occur on short timescales, potentially contributing to further field amplification or restructuring.

Following this early evolution, the magnetar magnetic field can become buried beneath the NS crust due to hypercritical fallback accretion \citep{chevalier1989neutron, art_dany, geppert1999submergence, mukherjee2017revisiting}. 
As large amounts of material accumulate on the stellar surface, the original dipole field is submerged into deeper layers, reducing the external field strength and suppressing observable magnetar signatures \citep{bernal2010hypercritical, Bernal2013, konar2017magnetic}. Once the hypercritical accretion phase ceases, the buried field gradually diffuses back to the surface \citep{haskell2008modelling,art_vigano, mukherjee2017revisiting}.

Numerical Hall–Ohm studies
\citep{art_vigano,TorresForne2016}\footnote{See also the discussion in
\citet{Bernal2013}.} show that the fractional burial depth,
$d_{\rm bur}/R_\star$, varies by at most $\sim20$\,\% when the seed dipole is changed from $10^{12}$ to $10^{14}$\,G; the process is thus
largely insensitive to $B_{0}$ when measured in relative rather
than absolute terms.
This was confirmed by two- and three-dimensional MHD simulations
\citep{fraija2014signatures, fraija2015hypercritical, fraija2018hypercritical}.  The delayed reemergence of the field can substantially influence the GRB afterglows, thereby delaying the onset of strong magnetar-driven emission. Observational and theoretical investigations suggest that millisecond magnetars linked to GRBs experience substantial magnetic field amplification, perhaps elucidating certain late-time emission features.

\begin{figure}
\centering
\includegraphics[width=0.95\linewidth]{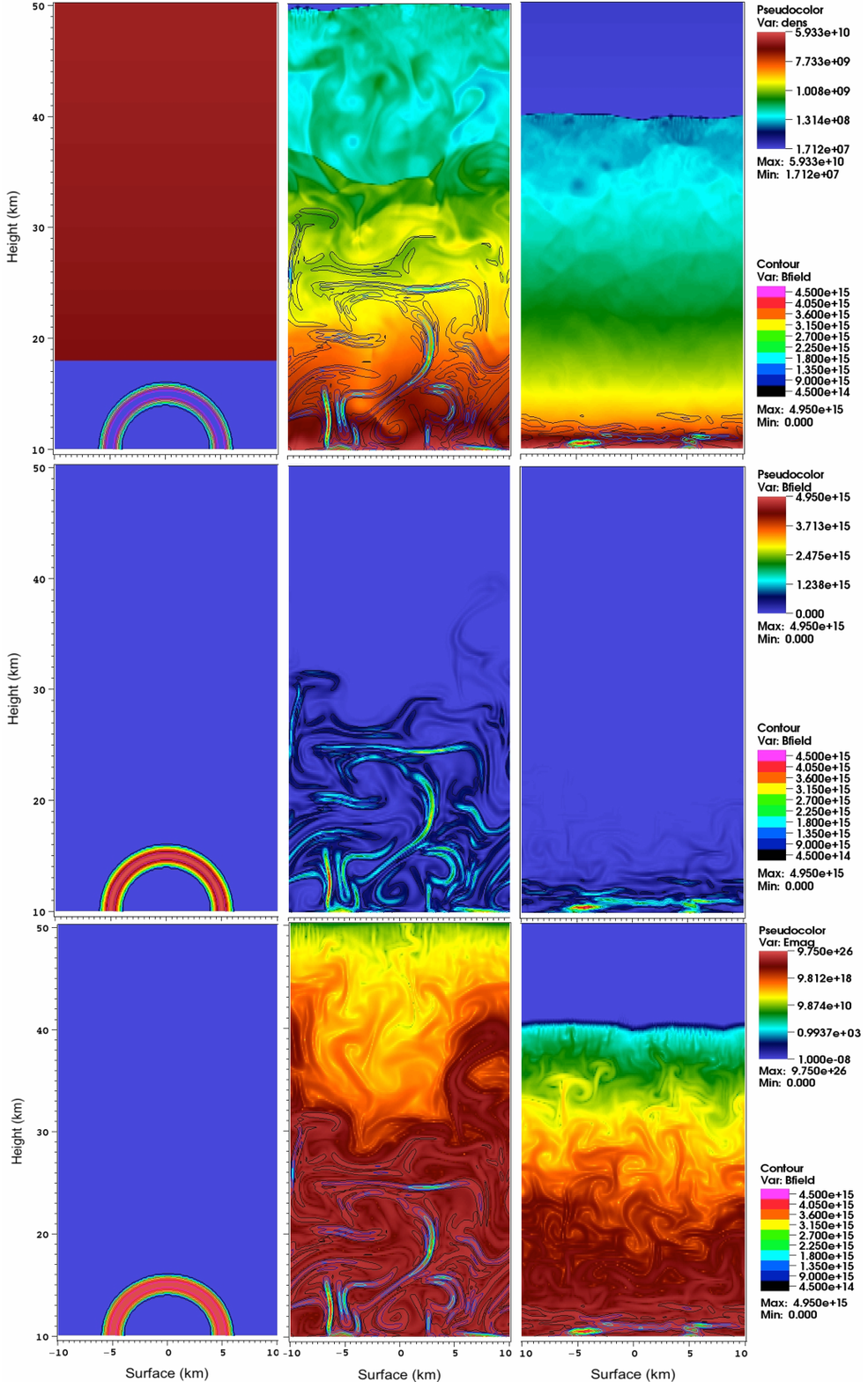}
\caption{Time evolution of the accretion process onto a magnetized NS. The top row displays density color maps of the accreted material with iso-contours representing the magnetic field loop configuration. The middle row shows magnetic field strength with superimposed magnetic iso-contours, and the bottom row illustrates magnetic energy maps at $t$ = 0, 10, and 100 ms, depicting the interaction between the accretion flow and the magnetic field. This model follows the approach outlined by \citet{Bernal2013}.}
\label{fig:bucles}
\end{figure}

The dynamics of this burial process are illustrated in Figure~\ref{fig:bucles}, which presents three representative time slices—$t$ = 0 ms, $t$ = 10 ms, and $t$ = 100 ms—capturing the evolution of the hypercritical accretion phase.  The upper panels exhibit the density of matter, illustrating how the fallback material progressively settles onto the NS surface, forming a newly accreted crust. At $t = 0\, {\rm ms}$, the stellar exterior remains unchanged, whereas by  $t = 10\,{\rm ms}$ and  $t = 100\,{\rm ms}$, dense layers have accumulated, pushing the magnetic field downward. The middle panels illustrate the magnetic field strength, with iso-contours indicating the compression and distortion of field lines as material deposition continues.  Initially, the dipole was prominent near the surface but became increasingly confined to deeper layers over time. Finally, the lower panels show the magnetic-energy density, where high-energy-density regions become localized beneath the new crust, further reducing the external manifestation of the magnetar-like field. These simulations performed using the FLASH code \citep{fryxell2000flash} confirm that hypercritical accretion naturally leads to field submergence, which may substantially postpone the appearance of strong magnetar-driven signals. This burial process can suppress the surface dipole field of the NS for years to centuries unless external diffusion mechanisms, such as crustal fractures, convective instabilities, or thermal changes, facilitate faster reemergence \citep{bernal2010hypercritical,Bernal2013, art_vigano, fraija2018hypercritical}. Meanwhile, the weak external field alters the detectability of the NS, minimizing dipole-driven pulsar activity at radio frequencies and modifying its X-ray and gamma-ray emissions \citep{art_dany, geppert1999submergence, mukherjee2017revisiting}. From a GRB phenomenology perspective, this delayed field exposure may explain specific late-time emission components, as magnetar spin-down luminosity could remain unobservable until the buried field resurfaces. The fate of this buried field—and the timescale of its reemergence—forms the central theme of the following subsection.

\subsection{On the Crustal Magnetic Field in Magnetars}

Millisecond magnetars are distinguished from ordinary NS by their exceptionally short spin periods and intense magnetic fields, reaching or exceeding 
$10^{14}$--$10^{15}\,\mathrm{G}$ at the surface \citep{1992ApJ...392L...9D, 1993ApJ...408..194T}. The large-scale dipole component often receives the majority of observational focus; however, the configuration of the crustal magnetic field is also crucial in influencing the star's evolution and emission characteristics \citep{art_dany, geppert1999submergence, haskell2008modelling, mukherjee2017revisiting}. In particular, the interplay between the ultra-strong magnetic field, rapid rotation, and the thermal and compositional profiles of the crust can drive a range of magneto-thermal and magneto-mechanical processes \citep{Perna2011AUM, art_vigano, konar2017magnetic}.  They are often enhanced in millisecond magnetars, in the GRB scenario, due to potential hypercritical accretion and intense heating shortly after birth \citep{chevalier1986pulsars, bernal2010hypercritical, metzger2011protomagnetar}.

The crust of a magnetar differs from that of an ordinary NS in two primary ways. First, the magnetic pressure, 
$P_B = B^2/(8\pi)$, becomes significant when surface fields exceed 
$10^{14}\,\mathrm{G}$, potentially modifying the local matter distribution and inducing crustal fractures \citep{gourgouliatos2014hall, haskell2008modelling}. In the millisecond magnetar scenario, rapid spin-down further amplifies these stresses, enabling partial or periodic breakage of the crust, which can in turn alter the geometry of the buried magnetic field. In addition, hyperaccretion and the associated nuclear processes in the newly formed crust generate substantial internal heating \citep{geppert1999submergence, bernal2010hypercritical, Bernal2013, konar2017magnetic}. This leads to strong thermal gradients that affect temperature-dependent electrical conductivities and impurity content \citep{cumming2006long}, both of which control the efficiency of current diffusion and rearrangement within the crust, thereby governing the timescale over which a large-scale dipole field can reemerge. 

Just as in ordinary NS crusts, the magnetic field in magnetars evolves according to the induction equation, which includes both Ohmic diffusion and the Hall effect \citep{shalybkov1995ambipolar, cumming2004magnetic, mukherjee2017revisiting, skiathas2024combined}. Written schematically,

\begin{equation}
\label{eq:HallOhmic}
\frac{\partial \mathbf{B}}{\partial t}
= -\,\nabla \times \Bigl[
   \underbrace{\bigl(\frac{\nabla \times \mathbf{B}}{n_e e}\bigr) \times \mathbf{B}}_{\text{Hall term}}
  + \underbrace{\frac{c^2}{4\pi\sigma}\,\nabla \times \mathbf{B}}_{\text{Ohmic term}}
\Bigr],
\end{equation}
\\
\noindent where $n_e$ is the electron number density, $e$ the elementary charge, $\sigma$ the electrical conductivity, and $c$ the speed of light. The Hall term is non-dissipative, but can redistribute magnetic energy by generating small-scale structures that then dissipate resistively through the Ohmic term \citep{pons2007magnetic, art_vigano}.

For standard field strengths ($B \lesssim 10^{13}$\,G), Hall drift is negligible on $\sim$decade timescales. However, for magnetar-level fields ($B \gtrsim 10^{14}$\,G), Hall dynamics can dominate the early field evolution \citep{kojima2012magnetic}.

\medskip
\noindent\textbf{Diffusion-limited evolution:}  
Using the conductivity tables of \citet{ChamelHaensel2008} at a melting temperature $T \simeq 5 \times 10^8$\,K, the Ohmic timescale is obtained:

\begin{equation}
\tau_{\rm Ohm} \simeq 4 \times 10^3\,\text{yr} \,
\Bigl( \frac{\sigma}{10^{24}\,\text{s}^{-1}} \Bigr)
\Bigl( \frac{L}{0.3\,\text{km}} \Bigr)^2,
\tag{2}
\end{equation}
\\
\noindent where $L$ is the depth of the submerged field. In cold, unperturbed NS crusts, this yields $\tau_{\rm Ohm} \gtrsim 10^4$--$10^6$\,yr. However, in hot, accretion-modified crusts---as expected in newborn millisecond magnetars---thermal and compositional effects can reduce $\sigma$ by several orders of magnitude \citep{geppert1999submergence, art_ho}. Moreover, since $\tau_{\rm Ohm} \propto L^2$, a shallow burial depth $L \sim 30$--$50$\,m (e.g., for $\Delta M \approx 10^{-4}\,M_{\odot}$) can shorten the Ohmic time to $\tau_{\rm Ohm} \lesssim 10^2$\,yr, even in the absence of additional transport processes.

\medskip
\noindent\textbf{Hall acceleration and optimistic scenarios:}  
Hall drift can further reduce the timescale via the magnetic Reynolds number,
$\mathcal{R}_{\rm H} = cB / 4\pi e n_e L$, yielding:

\begin{equation}
\tau_{\rm Hall} \approx 3\text{--}5\,\text{yr} \,
\Bigl( \frac{B}{3\times10^{15}\,\text{G}} \Bigr)^{-1}
\Bigl( \frac{n_e}{10^{31}\,\text{cm}^{-3}} \Bigr)
\Bigl( \frac{L}{0.3\,\text{km}} \Bigr).
\tag{3}
\end{equation}
\\
\noindent Thus, a diffusion time as short as $\tau_B \sim 3$\,yr is plausible only under extreme conditions:  
(i) an ultra-strong magnetic field ($B \gtrsim 3\times10^{15}$\,G);  
(ii) a burial layer confined to $L \lesssim 10^2$\,m; and  
(iii) efficient non-linear transport.  
For $B \lesssim 10^{15}$\,G, a more typical value is $\tau_B = 30$--$40$\,yr.

\medskip
\noindent\textbf{Patchy emergence and crustal flow:}  
Three-dimensional simulations show that Hall cascades, coupled to plastic or thermoplastic flow, can generate localized "hot spots" of magnetic re-emergence on month-to-year timescales \citep{LiBeloborodov2015, GourgouliatosCumming2014}. Since GRB afterglow spectra are dominated by compact emitting regions, such patchy, fast emergence may suffice to produce the observed hard X-ray excess, even if the global field resurfaces over longer timescales.

\medskip
\noindent\textbf{Thermal feedback:}  
Multidimensional simulations that couple magnetic and thermal evolution find that lateral heat transport into Hall-active regions can increase electron mobility and accelerate diffusion \citep{Aguilera2008, Vigano2013}. This "thermal-Hall" coupling may reduce the effective timescale by a factor of two to three. Nevertheless, achieving $\tau_B \simeq 3$\,yr still requires extreme field strengths or shallow burial. For $B \lesssim 10^{15}$\,G, the fiducial $\tau_B = 30$--$40$\,yr scenario remains the natural outcome, as illustrated in Figure~2.

\medskip
In summary, while a $\tau_B \sim 3$\,yr timescale is theoretically possible, it represents the optimistic limit of the parameter space. Most realistic systems likely fall in the broader range $10$--$40$\,yr, with the fastest emergence restricted to localized patches rather than the global field structure.

In addition to electromagnetic processes, mechanical stresses can further alter the crustal field structure. In magnetars, sudden fractures occur when the magnetic pressure exceeds the local yield strength; this temporarily allows buried flux tubes to move
more freely \citep{levin2012dynamics}. Moreover, plastic flow in magnetized crustal layers has been proposed as a mechanism that drags submerged field lines and may promote reconnection events \citep{chugunov2010breaking,li2015plastic}, potentially releasing flares or late-time afterglow signatures on timescales of a few years. For a GRB-associated magnetar, the external dipole field is thought to be buried under hyperaccreted material, but the crust may remain sufficiently hot and stressed to permit rapid field reconfiguration. The combination of
\textbf{(i)} Hall drift, \textbf{(ii)} partial Ohmic diffusion, and \textbf{(iii)} crustal
fractures or flows can bring the buried magnetic flux back to the surface in few years
\citep{fraija2018hypercritical,Metzger2018Effects,suvorov2020recycled}. This reemergence amplifies the
magnetar spin-down luminosity, potentially driving renewed activity in the late-time GRB
afterglow \citep{troja2017a, hajela2022evidence}. Such a scenario helps reconcile short
reemergence timescales with observational evidence for delayed X-ray and radio emission
occurring years after certain GRB events. Hence, while the rapid magnetic field growth over a few years is inherently complex—often demanding sophisticated multi-dimensional simulations to capture turbulence, reconnection, and crustal fracturing--it is not beyond physical plausibility. Such rapid field intensification can occur under the extreme conditions present in millisecond magnetars (e.g., high magnetic pressure, intense thermal gradients, and potential plastic flow in the crust). At the same time, simplified analytical approaches can provide valuable insights and predictive power regarding key energetic and dynamical consequences, guiding more advanced computational models and observational strategies designed to detect  the late-time signatures of a reactivated magnetar inside GRBs.

\subsection{On the Magnetic Field Reemergence}

As highlighted above, once hypercritical accretion ceases, the buried magnetic field of a magnetar may diffuse back through the crust, eventually restoring a pronounced dipole component at the stellar surface. Early suggestions of this mechanism indicate that substantial amounts of fallback material can keep a NS effectively unmagnetized from centuries to millennia \citep{michel1994magnetic,art_dany,geppert1999submergence, mukherjee2017revisiting}. Nevertheless, some authors have argued that a newborn NS field might reemerge on shorter time scales, thereby amplifying $B$ from $10^{10}{-}10^{11}\,\mathrm{G}$ to $10^{15}{-}10^{16}\,\mathrm{G}$ \citep[e.g., see][]{art_blandford,art_ho,art_vigano,Shabaltas_2012,konar2017magnetic}. In such models, an accreted mass of $\sim 0.01 \, \text M_{\odot}$— corresponding to an accretion rate of $\sim 100 \, \text M_{\odot} \, \text{yr}^{-1}$ over a few hours can sufficiently submerge the external dipole. This rapid growth of the magnetic field may proceed via reconnection, crustal instabilities, Hall drift, or thermomagnetic processes \citep{haskell2008modelling,pons2011magnetars,Shabaltas_2012, mukherjee2017revisiting}, often requiring detailed numerical
treatments for full accuracy \citep{vigelius2009resistive}.  In our fiducial scenarios, we adopt a parametric approach that allows for an "early" reemergence (i.e., on a timescale of $\sim\!\mathrm{few\ years}$) while acknowledging that real systems may deviate substantially due to, among other factors, (i)~the total fallback mass—larger values bury
the field deeper and thus prolong diffusion, (ii)~the post-accretion temperature—higher
crustal temperatures may alter resistivity and expedite field growth, and (iii)~the
magnetic geometry—complex multipoles can restructure at varying rates depending on crustal
fractures, Hall drift, and reconnection episodes \citep{haskell2008modelling,pons2011magnetars}.

Consequently, while our simplified prescription captures one plausible pathway for rapid field enhancement, a more conservative scenario—mainly if the fallback mass is appreciable—could lead to a substantially delayed reemergence \citep{vigelius2009resistive}. Crucially, these variations do not invalidate the qualitative concept of a "late-time boost" to the surface field; instead, they primarily affect the epoch at which this boost becomes apparent.  In standard spin-down theory, the rotational energy loss of a NS is typically treated under the assumption of a constant dipole magnetic field. However, magnetic field reemergence can substantially alter the burst energetics in millisecond magnetars associated with GRBs. To account for this effect, we extend the conventional framework by introducing a parametric model for the early magnetic field evolution of the magnetar, with parameters tuned to scenarios in which rapid dipole reactivation is plausible. Inspired by earlier works employing analytical prescriptions for magnetic-field growth \citep{art_blandford,negreiros2015growth,Rogers_2016,bernal2024overall}, our approach yields simple, tractable equations that capture the star’s early physical evolution without relying on complex numerical simulations. This enables us to investigate how a reemerging magnetic field affects the spin-down power and the overall dynamical properties of newborn millisecond magnetars.

In the present work, we propose that the magnetic field $B(t)$ undergoes a controlled growth phase before reaching the characteristic strength of a millisecond magnetar. In this framework, the field evolution determines the structural constant $k$ in the magnetar’s rotational dynamics with angular velocity ${\rm \Omega}$. By allowing $B(t)$ to evolve from an initially suppressed state to a canonical magnetar-level field, our approach captures how intrinsic properties—such as the crust’s resistive and thermal conditions—can significantly affect the star’s spin-down evolution and total energy output. With these considerations, the equation of motion of the NS is given by,

\begin{equation}
\dot{\Omega}=-k(t)\Omega^{n},\quad {\rm with}\quad k(t)=\frac{B^{2}R^{6}\sin^{2}\alpha}{6Ic^{3}}f(t)\,.
\label{MotionEquation}
\end{equation}

\noindent In this formulation, $n$ is the braking index, 
$B$ the maximum magnetic field strength, and 
$R$ the magnetar radius, while 
$I$ denotes the moment of inertia (assumed constant), 
$c$ the speed of light, and 
$\alpha$ the angle between the rotation and magnetic axes. Our model does not account for time-dependent alignment or misalignment; accordingly, we treat the magnetic field as the component perpendicular to the stellar surface.
We adopt analytic functions $f(t)>0$ to describe a rapid, diffusion-driven increase in $B$ over a prescribed timescale, allowing us to derive expressions for key physical quantities of the nascent magnetar, including the characteristic age, spin-down luminosity, and spin period as the field evolves. In particular, we highlight how magnetic field growth influences the spin-down luminosity, thereby affecting the overall energetics of the associated GRB.

By integrating the motion equation \ref{MotionEquation}, one can determine the magnetar’s characteristic age—an important measure of its evolving dynamics:

\begin{equation}
\tau=\frac{1}{f(t)}\left[\tau_{0}+\intop_{0}^{t}f(t')dt'\right],
\label{luminos}
\end{equation}

\noindent where $\tau_{0}$ is the magnetar’s initial spin-down timescale.
In turn, the spin-down luminosity evolves according to

\begin{equation}
\dot{E}=\dot{E}_{0}f(t)\left[\frac{\tau}{\tau_{0}}f(t)\right]^{-\frac{(n+1)}{(n-1)}},
\end{equation}

\noindent where $\dot{E}_{0}$ denotes the initial spin-down luminosity and $n=3$ is the canonical braking index. Although we have not yet specified the family of growth functions $f(t)$ that we want to study, any $f(t)$ chosen must capture a significant field increase over a characteristic diffusion timescale $\tau_{B}$. Here, $\tau_{B}$ encompasses multiple physical processes--Ohmic diffusion, Hall drift, potential reconnection, and related instabilities—folded into a single adequate timeframe. Consequently, the function $f(t)$ should rise from an initially small value (i.e., $\epsilon = (B_0/B)^2 \ll 1$) to unity by $t>\tau_B$ bridging the transition from a weakly magnetized state to the canonical regime of a millisecond magnetar. Once $f(t)=1$, our equations revert to the standard spin-down behavior of an oblique dipole rotator. 

While our formulation leaves the initial spin-down timescale \(\tau_{0}\) and the initial
spin-down luminosity \(\dot{E}_{0}\) as free parameters, these can be constrained via a
combination of theoretical and observational approaches
\citep{Rogers_2016,bernal2024overall}.
From an observational standpoint, one may leverage the early-time X-ray or optical
plateau in GRB afterglows—believed to be powered by the nascent magnetar—to estimate the
energy injection rate shortly after formation. By fitting the light-curve data with a
magnetar-driven spin-down model, \(\dot{E}_{0}\) can be inferred as the luminosity scale
at $t \!\lesssim\! \tau_{0}$. Likewise, the observed break in the afterglow flux or
transition to a steeper decay might provide an estimate of \(\tau_{0}\), marking when the
magnetar’s rotational energy contribution declines significantly.
On the theoretical side, detailed MHD simulations of hyperaccreting NSs can estimate the available rotational kinetic energy upon birth, thus helping constrain
\(\dot{E}_{0}\) (see, e.g., \citealt{Rogers_2016} and references therein). These
simulations also clarify how fallback accretion influences the magnetar’s moment of
inertia and the effective radius \(R\), supporting more precise evaluations of
\(\tau_{0}\). Similar parameter-sensitivity studies (e.g., \citealt{art_blandford,negreiros2015growth,bernal2024overall}) show that if the newborn star spins near breakup
(\(\sim\!\mathrm{1\ ms}\) periods) and retains a magnetar-strength field
(\(\gtrsim\!10^{14}\,\mathrm{G}\)), then \(\dot{E}_{0}\) can easily reach
$\sim\!10^{44}{-}10^{47}\,\mathrm{erg\,s}^{-1}$, with \(\tau_{0}\) typically in the range
$10{-}10^3\,$s. Consequently, once \(\dot{E}_{0}\) and \(\tau_{0}\) are pinned down—even
with modest precision—one can apply our analytical framework to forecast how an evolving
field \(B(t)\) reshapes the magnetar’s spin-down and, by extension, the GRB afterglow
energetics.

In Table~\ref{tab:math_functions}, we present three well-behaved functional forms for $f(t)$— an exponential, a hyperbolic, and a power law— each parameterized by $\epsilon$ and $\tau_{B}$, and all satisfying the fundamental requirement of a rapid, physically motivated growth of the newborn magnetar’s dipole field. While these forms are phenomenological, each can be loosely associated with known physical processes in the crust or magnetosphere of a millisecond magnetar, allowing us to explore a range of plausible reemergence scenarios
\citep{art_blandford,Rogers_2016,bernal2024overall}:
(i) Exponential:  If resistive diffusion remains roughly constant in time and the crust is kept at a nearly uniform temperature, one might anticipate an exponential-like relaxation toward the final, magnetar-level dipole field. Such behavior recalls classical Ohmic decay or equilibrium
approaches inhomogeneous media \citep[e.g.,][]{geppert1999submergence, konar2017magnetic}, where the timescale for a single, near-constant diffusivity dominates field evolution. Physically, exponential $f_{\exp}(t)$ could approximate scenarios in which the thermal and compositional
profiles change only slowly relative to the crustal diffusion timescale;
(ii) Hyperbolic/Saturation:  In some crustal evolution regimes---particularly those featuring partial Hall drift, reconnection events, or other self-regulating feedbacks---the magnetic field may grow steadily until saturating at its surface strength \citep{haskell2008modelling}. Here, $f_{\rm hyp}(t)$
represents a gradual, monotonic rise that transitions smoothly from early submergence to a "plateau" near full reemergence. Such behavior might characterize situations where local
instabilities expedite field movement but taper off once the external flux nears its
equilibrium configuration;
(iii) Power law: 
Nonlinear processes, including crustal fracturing, can drive a series of sporadic,
reconfiguration events that, in aggregate, appear as a power-law (PL) growth for the external
field \citep{1992ApJ...392L...9D,pons2011magnetars}. In this viewpoint, each sudden fracture or
magnetic reconnection event advances the magnetic flux outward; when averaged over time,
the net growth exhibits a PL trend. This could be especially relevant if the star’s
crust is repeatedly stressed by differential rotation or is riddled with impurity
gradients.

We stress that these three functions serve only as illustrative \textit{parametric}
templates for $f(t)$. Real NS crusts likely experience a complex interplay of
Ohmic diffusion, Hall drift, reconnection, and plastic flow \citep{suvorov2020recycled}, thus
no single analytical form fully captures the multiscale nature of magnetic-field evolution.
Nonetheless, adopting these simplified functions provides a straightforward analytical tool
to examine how adjusting the diffusion timescale $\tau_{B}$ or the initial growth ratio
$\epsilon$ (i.e., $B_{0}/B \ll 1$) influences magnetar spin-down and the resulting
afterglow emission.  By relating $\tau_{B}$ to previously studied timescales for
magnetic burial and reemergence \citep[e.g.,][]{konar2002diamagnetic,10.1111/j.1365-2966.2004.07798.x,art_vigano, Bernal2013, bernal2024overall},
we constrain it to roughly the $1{-}100$\,yr range while acknowledging that the shorter end
($1{-}10$\,yr) is an optimistic scenario aligned with late afterglow data for events such as
GW170817/GRB 170817A \citep{hajela2022evidence}. Thus, although these three functional forms do
not replace sophisticated numerical simulations, they allow us to capture key features of the
field’s reemergence and systematically explore how a time-varying magnetic field can
influence GRB afterglows.

As extensively discussed, the evolution of strongly magnetized NSs under early
hyperaccretion is intrinsically complex, involving factors that extend far beyond the conventional
rotation-powered pulsar (RPP) framework. Indeed, although canonical spin-down theory offers a foundation for
interpreting NS evolution, recent observations and both analytical and numerical studies emphasize
the potential for prompt magnetic field reemergence in newborn magnetars
\citep{art_vigano,Shabaltas_2012,Rogers_2016}, naturally explaining unusual features in
the light curves and spectra of certain GRBs when crustal diffusion, reconnection, or plastic flow
becomes significant \citep{haskell2008modelling, mukherjee2017revisiting}. 

\begin{figure}
\centering
\includegraphics[width=1\linewidth]{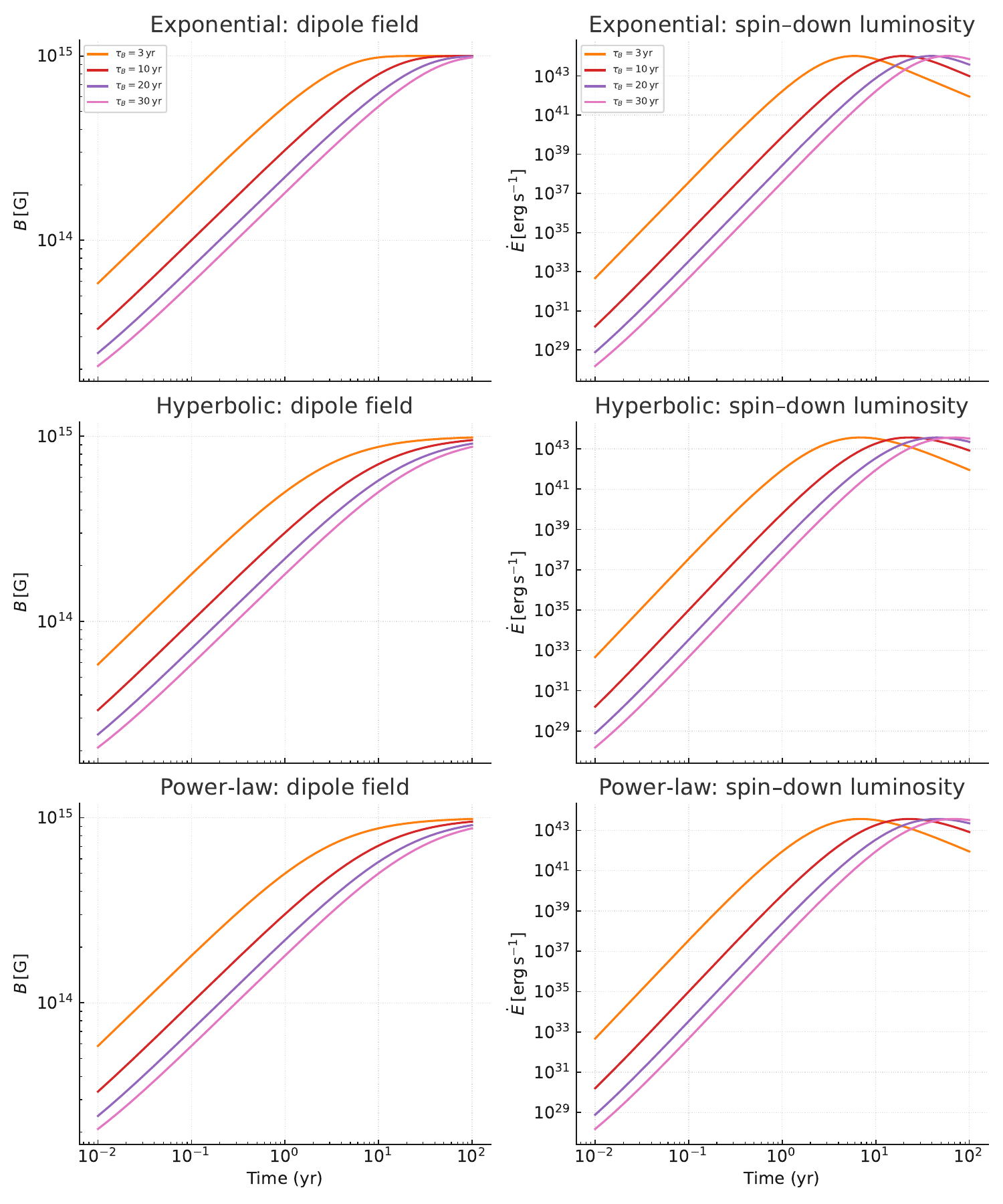}
\caption{
Magnetic-field growth and spin-down luminosity for three functional forms of the diffusion law $f(t)$. Rows (top to bottom) correspond to exponential, hyperbolic, and power-law prescriptions (with $\alpha = 1$ in the latter). 
Left-hand panels: evolution of the surface dipole field $B(t) = B_{\max} f(t)$, with $B_{\max} = 10^{15}$\,G.
Right-hand panels: rotational energy loss $\dot{E}(t)$ computed using Equation~(5), assuming a braking index $n = 3$.
Coloured curves show four diffusion timescales, $\tau_B = 3$, 10, 20, and 30\,yr, with the initial spin-down time set equal to the diffusion timescale ($\tau_B$). We adopt an initial suppression factor $\epsilon = 10^{-4}$ and a canonical unsuppressed luminosity $\dot{E}_0 = 10^{45}$\,erg\,s$^{-1}$. Each curve therefore starts at $\dot{E} \simeq \epsilon^5 \dot{E}_0$, rises as the magnetic field re-emerges, peaks near the canonical value once $f \to 1$, and subsequently decays as $\dot{E} \propto t^{-2}$.
}
\label{fig:magfield}
\end{figure}

Figure~\ref{fig:magfield} compares the three field--growth
prescriptions for four diffusion timescales,
$\tau_{B}=3,\,10,\,20,\,30$~yr, with the characteristic spin--down time
set equal to the corresponding diffusion time
($\tau_{B}$).  The left--hand panels show that for each
prescription,  the surface field $B(t)$ increases by four orders of magnitude ---from
an initially suppressed value of $\sim10^{11}\,\mathrm G$
($\epsilon=10^{-4}$) to the magnetar strength
$\gtrsim10^{15}\,\mathrm G$---within a few~$\tau_{B}$
\citep{geppert1999submergence, art_vigano, Shabaltas_2012,
Rogers_2016, bernal2024overall}.  
The right--hand panels plot the full spin--down luminosity
$\dot E(t)$ for each $\tau_{B}$.  Because the torque is initially
weakened by the buried field, every curve starts well below the
canonical power; as $f(t)$ approaches unity, the luminosity grows and
reaches a maximum essentially equal to the unsuppressed value,
$\dot E_{0}\simeq10^{45}\ {\rm erg\,s^{-1}}$.  After saturation the
evolution re--joins the standard dipole decay,
$\dot E\propto t^{-2}$.  These rise--peak--decay sequences are in line
with the Hall--Ohm simulations of crustal diffusion
\citep{vigelius2009resistive, haskell2008modelling, art_vigano,
Bernal2013} and with previous analytic studies of early field
amplification in millisecond magnetars
\citep{Rogers_2016, bernal2024overall}.  As emphasised by
\citet{2005ApJ...634L.165S}, a newborn magnetar is expected to possess
a spin period of a few~ms and a luminosity of order
$10^{44}$--$10^{45}\,{\rm erg\,s^{-1}}$; this figure makes
explicit how a buried--field episode can delay the appearance of that
canonical output by several to a few tens of years, depending on
$\tau_{B}$.

Crucially, spin-down luminosity is the primary quantity that dictates energy availability for late-time GRB afterglow (Sections 3 and 4), thus
tying the timing of this rapid magnetic field growth to potentially observable features in
multiwavelength light curves, highlighting that the transition from a buried to a strong dipole
component provides a necessary basis for understanding GRB afterglow dynamics and for our targeted
application to GW170817/GRB,170817A.

\section{Theoretical approach:  Dynamics of the GRB afterglow and Synchrotron Light Curves}\label{sec3:model}


The dynamics of the decelerated material are expected to lie in the nonrelativistic regime over a few years, and therefore, it is characterized by the Sedov–Taylor solution. In the GRB afterglow phase, the relativistic jets interact with the external medium, transferring a significant portion of their energy, resulting in the formation of two shock waves: a forward shock and a reverse shock \citep[e.g., see][]{1995ApJ...455L.143S,2000ApJ...545..807K}.   We focus on the dynamics of forward shocks, excluding reverse shocks, as electrons accelerated in reverse shocks produce transient emissions. In contrast, we aim to characterize emissions that persist across periods of years. We consider the dynamics of the afterglow model with constant density (${\rm n_s}$) because the circumburst environment at a shock-decelerated radius of the nonrelativistic material is expected to be homogeneous at a timescale of years.   Note that, at this timescale,  the shock-accelerated electrons lie in the slow-cooling regime instead of the fast-cooling regime. Once the afterglow transitions into the slow-cooling phase, the adiabatic scenario could model the hydrodynamic development \citep[e.g., see][]{2000ApJ...529..151M}. Therefore, we only consider the adiabatic regime instead of the radiative one. Therefore, the synchrotron light curves in the fast-cooling regime are included for completeness, although they are outside the issues examined in the present paper.

In the deceleration phase, the ejected mass develops a velocity gradient, with the matter at the front moving more rapidly than that at the rear \citep{2000ApJ...535L..33S}. \cite{2001ApJ...551..946T} simulated the deceleration at sub-relativistic velocities of the expelled mass by the circumstellar medium, finding that for a polytropic index $n_p=3$, the isotropic-equivalent kinetic energy can be described as a PL velocity distribution represented by $E_{\rm \beta} (\geq \beta)= \tilde{E}\,\beta^{-\alpha}$, where $\tilde{E}$ is the fiducial energy and $\alpha$ lies in the range $[3 -  5.2]$ \citep{2001ApJ...551..946T}. This description corresponds to the dynamics of a quasi-spherical material that decelerated in a homogeneous environment. In the case of a non-spherical ejecta, the dynamics of the afterglow will be described considering $\alpha=0$.

\subsection{Synchrotron Light Curves before B-reemergence}

We consider that the electrons population with a spectral index $p$ and minimum Lorentz factor ($\gamma_{\rm m}$) accelerated by the external forward shocks can be characterized by a simple PL energy distribution $\frac{dn}{d\gamma_e}\propto \gamma_e^{-p}$ for $\gamma_{\rm e}\geq \gamma_{\rm m}$. We use prime and unprimed quantities for the comoving and observer frames, respectively. In the early phase of expansion, the ejecta mass remains unaffected by the circumstellar medium \citep{2014MNRAS.439..757G}, resulting in a constant velocity ($\beta \propto t^0$) and a radius that grows as $r \propto t$. During the free coasting phase, the magnetic field in the post-shock region varies as {\small $B'\propto \,t^0$}, and the Lorentz factors of the low-energy electrons and those high-energy electrons that cooled down by synchrotron mechanism as $\gamma_{\rm m}\propto t^0$ and $\gamma_{\rm m}\propto t^{-1}$, respectively. The corresponding synchrotron spectral breaks vary as $\nu_{\rm m}\propto t^0$ and  $\nu_{\rm c}\propto t^{-2}$, respectively, and the maximum flux of synchrotron emission as  $F_{\rm \nu, max}\propto t^3$. On the one hand, for $\gamma_{\rm c} < \gamma_{\rm m}$ the synchrotron light curve is in the regime of fast cooling, and on the other hand, this light curve is in the slow cooling. At this early phase, the synchrotron light curves during fast-cooling regime evolve as  $F_{\rm \nu}\propto t^{\frac{11}{3}}$ for $\nu<\nu_{\rm c}$, $\propto t^2$ for $\nu_{\rm c}<\nu<\nu_{\rm m}$ and $\propto t^2$ for $\nu_{\rm m}<\nu$. During the slow-cooling regime, the synchrotron light curves vary as $F_{\rm \nu}\propto t^3$ for $\nu<\nu_{\rm m}$, $\propto t^3$ for $\nu_{\rm m}<\nu<\nu_{\rm c}$ and $\propto t^2$ for $\nu_{\rm c}<\nu$.

In the deceleration phase, the velocity and the radius of the blast wave develop as $\beta \propto t^{-\frac{3}{5}}$ and $r \propto t^{\frac{2}{5}}$, respectively.  Similarly, the magnetic field in the post-shock region varies as {\small $B'\propto \,t^{-\frac{3}{5}}$}. It is worth mentioning that while the shock is occurring, a fraction of the overall energy is continually being transferred to accelerate the electron population and increase the strength of the magnetic field through microphysical parameters $\varepsilon_e$ and $\varepsilon_B$, respectively.  During this phase, the minimum and cooling Lorentz factors of the electron population are

{\small
\bary\label{gamma_dec}
\gamma_{\rm m}&=&\,\left(\frac{1+z}{1.022}\right)^{\frac{6}{\alpha+5}}\, \left(\frac{p-2}{p-1}\right) \,\varepsilon_{\rm e,-1}\, n_{\rm s}^{-\frac{2}{\alpha+5}}\,\tilde{E}_{51}^{\frac{2}{\alpha+5}}\, t_7^{-\frac{6}{\alpha+5}}\cr
\gamma_{\rm c}&=&\left(\frac{1+z}{1.022}\right)^{-\frac{1-\alpha}{\alpha+5}}\, (1+Y)^{-1} \varepsilon^{-1}_{\rm B,-2}\, n_{\rm s}^{-\frac{\alpha+3}{\alpha+5}}\,\tilde{E}_{51}^{-\frac{2}{\alpha+5}}\, t_7^{\frac{1-\alpha}{\alpha+5}}\,,
\eary
}

respectively, where $Y$ is the Compton parameter \citep[e.g., see][]{2023ApJ...958..126F}, and $\varepsilon_{\rm e}$ and $\varepsilon_{\rm B}$ correspond to the fraction of energy required to accelerate electrons and amplify the magnetic field, respectively. Hereafter, the convention $Q_{\rm x}=Q/10^{\rm x}$ in cgs units is adopted throughout this manuscript. It is worth noting that when the decelerated material enters the deep Newtonian regime, the minimum Lorentz factor becomes $\gamma_m\simeq 2$ \citep{2013ApJ...778..107S,2020arXiv200413028M}.  The electron Lorentz factors produce spectral breaks in the synchrotron light curves. Therefore, the characteristic and cooling spectral breaks are

{\small
\bary\label{nu_syn_de_v1}
\nu_{\rm m}&=&\,\left(\frac{1+z}{1.022}\right)^{\frac{20-2\alpha }{2(\alpha+5)}}\,\left(\frac{p-2}{p-1}\right)^2\, \varepsilon^2_{\rm e,-1}\,\varepsilon^\frac12_{\rm B,-2}\,  n_{\rm s}^{\frac{\alpha-5}{2(\alpha+5)}}\, \tilde{E}_{51}^{\frac{5}{(\alpha+5)}}\,t_7^{-\frac{15}{(\alpha+5)}}\cr
\nu_{\rm c}&=&\,\left(\frac{1+z}{1.022}\right)^{-\frac{4-\alpha}{(\alpha +5)}}\, \epsilon^{-\frac32}_{\rm B,-2}\, (1+Y)^{-2} \, n_{\rm s}^{-\frac{3(\alpha+3)}{2(\alpha+5)}}\tilde{E}_{51}^{-\frac{3}{(\alpha+5)}}\,t_7^{-\frac{2\alpha+1}{(\alpha+5)}}\,,
\eary
}

 respectively.   The spectral breaks of the self-absorbed synchrotron emission are
 {\small
\bary\label{nua_syn_de}
\nu_{\rm a,1}&=&  \left(\frac{1+z}{1.022}\right)^{-\frac{55+8\alpha}{5(\alpha+5)}} \left(\frac{p-2}{p-1}\right)^{-1} \varepsilon_{\rm e,-1}^{-1} \varepsilon_{\rm B,-2}^{\frac15}\, n_{\rm s}^{\frac{25+4\alpha}{5(\alpha+5)}}\tilde{E}_{51}^{-\frac{5}{5(\alpha+5)}}  t_6^{\frac{30+3\alpha}{5(\alpha+5)}}\,\cr
\nu_{\rm a,2}&=&  \left(\frac{1+z}{1.022}\right)^{\frac{-2p(-10+\alpha)-12(5+\alpha)}{2(p+4)(\alpha+5)}}\left(\frac{p-2}{p-1}\right)^{\frac{2(p-1)}{p+4}} \varepsilon_{\rm e,-1}^{\frac{2(p-1)}{p+4}}\, n_{\rm s}^{\frac{p(\alpha-5)+6(\alpha+5)}{2(p+4)(\alpha+5)}} \varepsilon_{\rm B,-2}^{\frac{p+2}{2(p+4)}}\,\tilde{E}_{51}^{\frac{10p}{2(p+4)(\alpha+5)}}\, t_6^{-\frac{15p-2(5+\alpha)}{(p+4)(\alpha+5)}},\cr
\nu_{\rm a, 3}&=&  \left(\frac{1+z}{1.022}\right)^{-\frac{20+13\alpha}{5(\alpha+5)}}\,(1+Y)\, \varepsilon_{\rm B,-2}^{\frac65}\, n_{\rm s}^{\frac{3(3\alpha+10)}{ 5(\alpha+5)}} \tilde{E}_{51}^{\frac{15}{5(\alpha+5)}} t_6^{\frac{8\alpha-5}{5(\alpha+5)}}.
\eary
}

Taking into account that the peak spectral power evolves as $P_{\rm \nu, max} \propto\,t^{-\frac{3}{(\alpha+5)}}$ and the number of swept-up electrons in the post-shock develops as $N_{\rm e}\propto  t^\frac{6+3\alpha}{\alpha+5}$,  the spectral peak flux density becomes
{\small
\bary\label{f_syn_de}
F_{\rm \nu,max}&=&\,\left(\frac{1+z}{1.022}\right)^{\frac{2-2\alpha}{(\alpha+5)}}\, \varepsilon^{\frac12}_{\rm B,-2}\, d_{\rm z,26.5}^{-2}\, n_{\rm s}^{\frac{3\alpha+7}{2(\alpha+5)}}\,  \tilde{E}_{51}^{\frac{4}{(\alpha+5)}}\,t_7^{\frac{3 + 3\alpha}{(\alpha+5)}}.\,\,\,\,\,
\eary
}
 Using the synchrotron break frequencies (eq.\ref{nu_syn_de_v1}) and the spectral peak flux density (eq.\ref{f_syn_de}),  we derive in Table \ref{Table2} the evolution in the adiabatic regime of spectral and temporal PL indexes associated with the synchrotron light curve and its respective closure relations ($F_\nu \propto t^{-\alpha_c}\nu^{-\beta}$) without the reemergence of magnetic field for the cooling conditions $\nu_{\rm a,3}\leq \nu_{\rm c}\leq \nu_{\rm m}$, $\nu_{\rm a,1}\leq \nu_{\rm m}\leq \nu_{\rm c}$ and $\nu_{\rm m}\leq \nu_{\rm a, 2}\leq \nu_{\rm c}$. This table shows (from left to right) the frequency in each cooling condition, the PL of spectral ($\beta$), and temporal ($\alpha_c$) indexes, and the closure relation ($\alpha_c(\beta_c)$). It is worth noting that when $\alpha=0$, the evolution of the synchrotron light curves and the closure relations correspond to the dynamics of a nonspherical material that decelerates in a homogeneous environment.

\subsection{Light Curves after B-reemergence}

In the case of the exponential function type one ($f_1(t)$) given in Table \ref{tab:math_functions}, we can derive analytical synchrotron light curves with $\varepsilon\ll 1$ and $t\ll \tau_B$. In this case, the spin-down luminosity becomes

\bary\label{Lin}
\dot{E}(t)= \dot{E}_{\rm 0}\, \left(\frac{\tau}{\tau_0}\right)^{-\frac{n+1}{n-1}} \left(\frac{t}{\tau_{\rm B}} \right)^{-\frac{2}{n-1}}\,.
\eary
%
%
During the deceleration phase,  the post-shock magnetic field evolves as $B'\propto \,t^{-\frac{2n}{\alpha+5}}$, and the minimum and cooling Lorentz factors of the electron population are
{\small
\bary\label{gamma_dec_B}
\gamma_{\rm m}&=&\,\left(\frac{1+z}{1.022}\right)^{\frac{6}{\alpha+5}}\,\left(\frac{p-2}{p-1}\right)\, \varepsilon_{\rm e,-1}\, n_{\rm s}^{-\frac{2}{\alpha+5}}\,\tilde{E}_{51}^{\frac{2}{\alpha+5}}\, t_7^{-\frac{4n}{(n-1)(\alpha+5)}}\cr
\gamma_{\rm c}&=&\left(\frac{1+z}{1.022}\right)^{-\frac{1-\alpha}{\alpha+5}}\, (1+Y)^{-1} \varepsilon^{-1}_{\rm B,-2}\, n_{\rm s}^{-\frac{\alpha+3}{\alpha+5}}\, \tilde{E}_{51}^{-\frac{2}{\alpha+5}}\, t_7^{\frac{5+\alpha-n(\alpha+1)}{(n-1)(\alpha+5)}}\,,
\eary
}
respectively.  The corresponding synchrotron spectral breaks and the spectral peak flux density can be written as 
{\small
\bary\label{nu_syn_de}
\nu_{\rm m}&=&\,\left(\frac{1+z}{1.022}\right)^{\frac{10-\alpha }{\alpha+5}}\,\left(\frac{p-2}{p-1}\right)^{2} \varepsilon^2_{\rm e,-1}\,\varepsilon^\frac12_{\rm B,-2}\,  n_{\rm s}^{\frac{\alpha-5}{2(\alpha+5)}}\, \tilde{E}_{51}^{\frac{5}{\alpha+5}}\,t_7^{-\frac{10n}{(n-1)(\alpha+5)}}\cr
\nu_{\rm c}&=&\,\left(\frac{1+z}{1.022}\right)^{-\frac{4-\alpha }{\alpha +5}}\, \varepsilon^{-\frac32}_{\rm B,-2}\, (1+Y)^{-2} \, n_{\rm s}^{-\frac{3(\alpha+3)}{2(\alpha+5)}}\tilde{E}_{51}^{-\frac{3}{\alpha+5}}\,t_7^{\frac{2[5-2n-\alpha(n-1)]}{(n-1)(\alpha+5)}}\,\cr
F_{\rm \nu,max}&=& \,\left(\frac{1+z}{1.022}\right)^{\frac{2(1-\alpha)}{\alpha+5}}\, \varepsilon^{\frac12}_{\rm B,-2}\, d_{\rm z,26.5}^{-2}\, n_{\rm s}^{\frac{3\alpha+7}{2(\alpha+5)}}\,  \tilde{E}_{51}^{\frac{4}{\alpha+5}}\,t_7^{-\frac{15-7n+3\alpha(1-n)}{(n-1)(\alpha+5)}},\,\,\,\,\,
\eary
}
respectively.   In the self-absorption regime, the synchrotron break frequencies are 
{\small
\bary\label{nua_syn_de_B}
\nu_{\rm a,1}&=&  \left(\frac{1+z}{1.022}\right)^{-\frac{55+8\alpha}{5(\alpha+5)}} \left(\frac{p-2}{p-1}\right)^{-1} \varepsilon_{\rm e,-1}^{-1} \varepsilon_{\rm B,-2}^{\frac15}\, n_{\rm s}^{\frac{25+4\alpha}{5(\alpha+5)}}\tilde{E}_{51}^{-\frac{1}{\alpha+5}}  t_6^{-\frac{5(3-5n)-3\alpha(n-1)}{5(n-1)(\alpha+5)}}\,\cr
\nu_{\rm a,2}&=&  \left(\frac{1+z}{1.022}\right)^{-\frac{p(\alpha-10)+6(5+\alpha)}{(p+4)(\alpha+5)}}\left(\frac{p-2}{p-1}\right)^{\frac{2(p-1)}{p+4}} \varepsilon_{\rm e,-1}^{\frac{2(p-1)}{p+4}}\, n_{\rm s}^{\frac{p(\alpha-5)+6(\alpha+5)}{2(p+4)(\alpha+5)}} \varepsilon_{\rm B,-2}^{\frac{p+2}{2(p+4)}}\,\tilde{E}_{51}^{\frac{5p}{(p+4)(\alpha+5)}}\, t_6^{-\frac{2[5+\alpha-n(5+\alpha-5p)]}{(n-1)(p+4)(\alpha+5)}},\cr
\nu_{\rm a, 3}&=&  \left(\frac{1+z}{1.022}\right)^{-\frac{20+13\alpha}{5(\alpha+5)}}\,(1+Y)\, \varepsilon_{\rm B,-2}^{\frac65}\, n_{\rm s}^{\frac{3(3\alpha+10)}{ 5(\alpha+5)}} \tilde{E}_{51}^{\frac{3}{\alpha+5}} t_6^{-\frac{2[5(4-n)+4\alpha(1-n)]}{5(n-1)(\alpha+5)}}.
\eary
}

Using the synchrotron break frequencies and and the spectral peak flux density (eqs.~\ref{nu_syn_de}),  we derive in Table \ref{Table3} the evolution in the adiabatic regime of spectral and temporal PL indexes associated with the synchrotron light curve and its respective closure relations  ($F_\nu \propto t^{-\alpha_c}\nu^{-\beta}$) with the reemergence of magnetic field for the cooling conditions $\nu_{\rm a,3}\leq \nu_{\rm c}\leq \nu_{\rm m}$, $\nu_{\rm a,1}\leq \nu_{\rm m}\leq \nu_{\rm c}$ and $\nu_{\rm m}\leq \nu_{\rm a, 2}\leq \nu_{\rm c}$.  From left to right, this table exhibits cooling condition, the PL indexes of spectral ($\beta$), the temporal ($\alpha_c$), and closure relation  ($\alpha_c(\beta)$). Again, as the PL index of the velocity distribution becomes $\alpha=0$, the evolution of synchrotron flux corresponds to the dynamics of the homogeneous afterglow of a nonspherical material.

\subsection{Analysis of the light curves}

\begin{figure}
\centering
\includegraphics[width=0.85\linewidth]{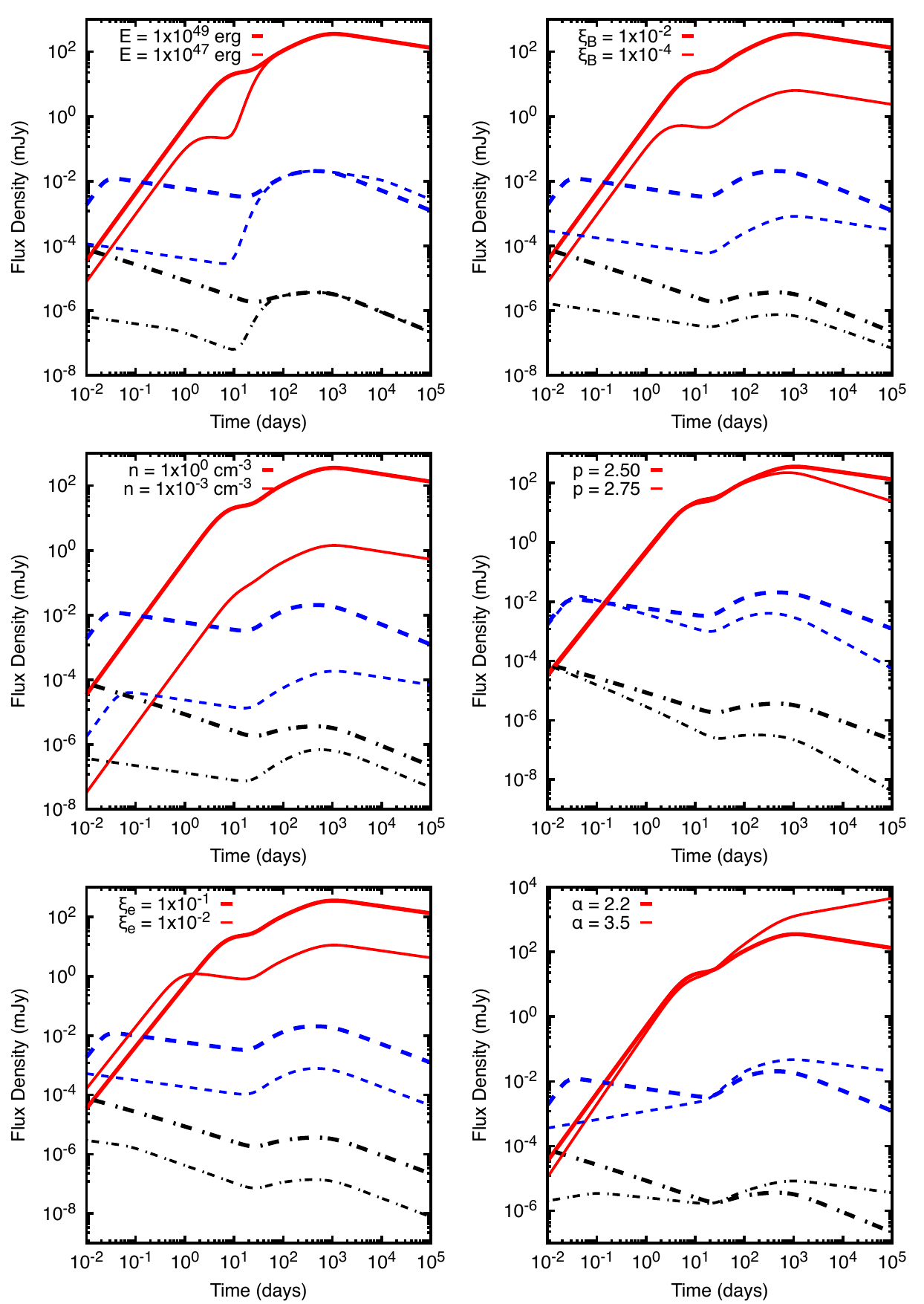}
\caption{Synchrotron light curves at X-ray (black), optical (blue) and radio (red) bands generated  by the deceleration of the non-relativistic material in the external environment. The light curves of X-ray, optical and radio bands are estimated at 1 keV, 1 eV and 10 GHz, respectively. The parameters used for the emergence of B-field model are: $\tau_{B} = 3\, {\rm year}$, $\epsilon = 10^{-4}$, $\tau_{0} = 2.5\, {\rm year}$, $\dot{E_{0}} = 4.5\times 10^{43} \ {\rm erg/s}$. Each panel shows the variation of one parameter sequentially in our  afterglow scenario, but keeping others fixed.
In general these values are: E $=10^{49} \ {\rm erg}$, $n_{\rm s} = 1 \ {\rm cm^{-3}}$, $\xi_{e} = 10^{-1}$, $\xi_{B} = 10^{-2}$, $p = 2.5$, $\alpha = 2.2$, $\theta_j=4.6\,{\rm deg}$ and $\Delta \theta=15.0\,{\rm deg}$. These parameter values are reported in Table \ref{tab:par_values}.}
\label{fig:AfterglowVar}
\end{figure}

In Figure~\ref{fig:AfterglowVar}, we present the variations in physical parameters related to the afterglow emission across six panels, each representing a different parameter change.   The colors represent different energy bands; in red, we have X-rays; in blue, we have optical; and in black, we have radio. In the upper left panel, we vary the kinetic energy between $\tilde{E}=10^{49}\rm erg$ and $\tilde{E}=10^{49}~\rm erg$. As expected, increasing the total kinetic energy results in a higher overall flux for all wavelengths. The difference in flux is particularly pronounced in the optical and radio bands at early times, where the difference reaches around two orders of magnitude. We also note that the peak at $\sim10^3~\rm days$ is not as pronounced in the higher energy case. Furthermore, both cases seem to match at late times. As the shock expands into the circumburst medium, the kinetic energy gets distributed over an increasingly larger volume, and the shock decelerates.  This deceleration causes the initial difference in total energy between the high-energy and low-energy cases to become less significant.   In the middle left panel, we change the density of the surrounding medium and compare between $n_{\rm s}=1~\mathrm{cm}^{-3}$ and $n_{\rm s}=10^{-3}~\mathrm{cm}^{-3}$. As the density increases, the flux also does, especially in the X-ray band, where the higher-density medium produces a more luminous emission that peaks earlier.    The optical flux follows a similar trend, albeit not so pronounced. Contrastingly, late-time radio emission is less sensitive to density variations, with a less pronounced difference in peak flux. Higher densities produce to faster and brighter emissions, especially in high-energy bands such as X-rays.   The lower left panel shows variation in the electron energy fraction from $\varepsilon_{\rm e}=10^{-2}$ to $\varepsilon_{\rm e}=10^{-1}$. This leads to substantial changes in the behavior of the X-ray flux. At first, the flux from $\varepsilon_{\rm e}=10^{-2}$ dominates that from $\varepsilon_{\rm e}=10^{-1}$, but at $t\sim1~\rm day$ the flux from the lower fraction reaches its first peak while the flux from the higher fraction continues to rise and overcomes it.

The optical and radio bands are also affected, although the difference between both cases remains constant throughout the evolution. Also, in the lower energy bands, the microphysical parameter $\varepsilon_{\rm e}=10^{-1}$ case dominates throughout. We can make similar observations from the upper right panel, which varies the magnetic field energy fraction $\varepsilon_{\rm B}$.   The higher fraction $\varepsilon_{\rm B}=10^{-2}$ results in a more luminous afterglow, as it enhances synchrotron radiation. The middle right panel examines the variation in the electron spectral index $p$. There is no substantial variation between the light curves at early times, as these cooling regimes are independent of this index.    Later, a smaller spectral index generally leads to a more considerable flux. In this case, the lower energies are more sensitive to this difference, and variations can be observed early on at $t\sim10^{-1}~\rm days$, while in the X-ray band only until $t\sim10^{3}~\rm days$. Finally, the lower right panel explores the parameter $\alpha$. A lower value of $\alpha$ results in a sharper rise and a sharper decline, producing a higher flux at the beginning of the study but a lower flux at later times across all wavelengths.

\begin{figure}
\centering
\includegraphics[width=0.9\linewidth]{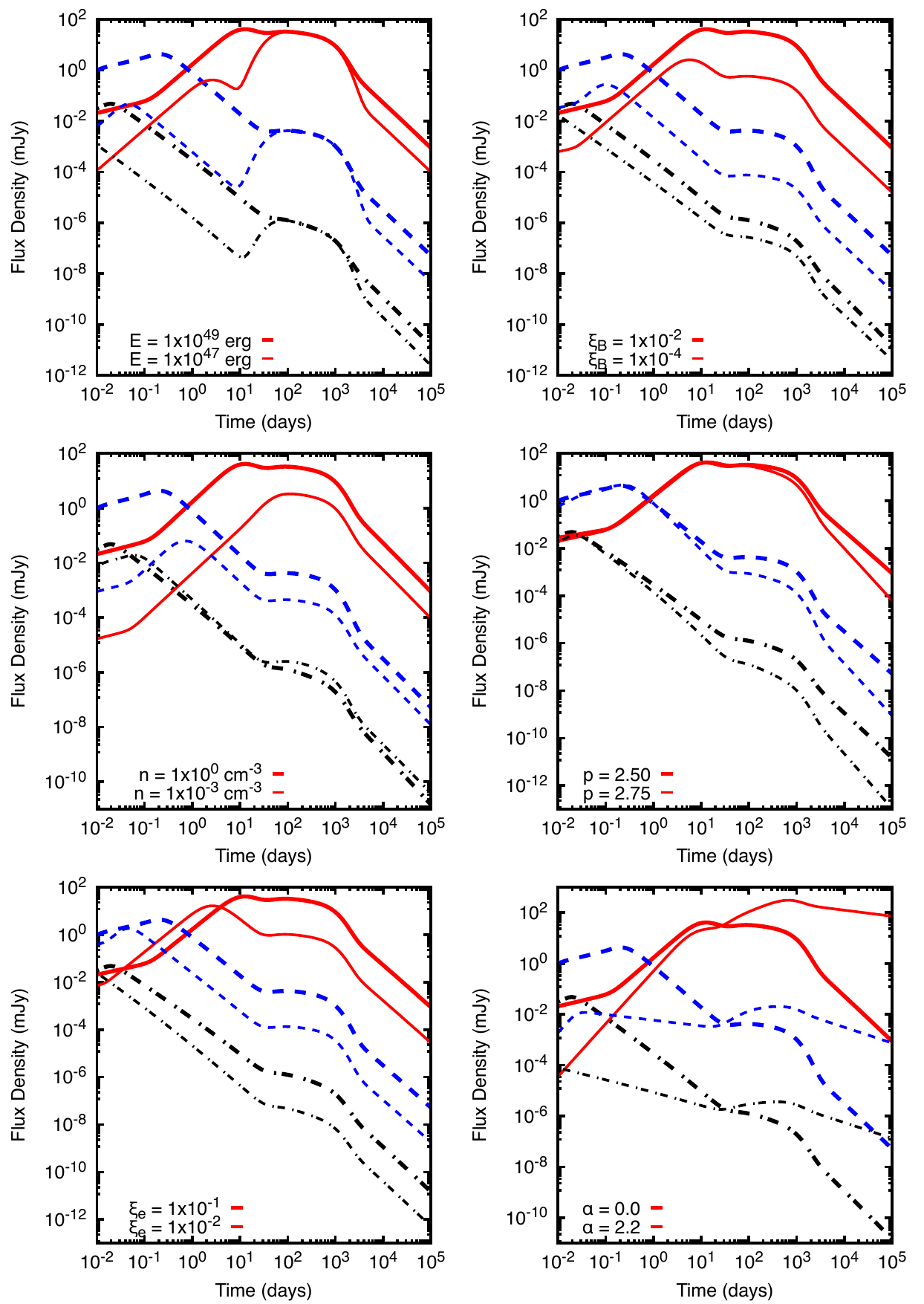}
\caption{The same as Figure \ref{fig:AfterglowVar}, but for $\alpha = 0.0$.}
\label{fig:AfterglowVarRel}
\end{figure}

In Figure~\ref{fig:AfterglowVarRel}, the primary difference from Figure ~\ref{fig:AfterglowVar} is the parameter $\alpha$ variation, where $\alpha=0.0$ is used instead of $\alpha=2.2$.  For $\alpha=0.0$, the dynamics and the synchrotron light curves correspond to those derived for a jet (e.g., top-hat jet) instead of a quasi-spherical material (distinct value of $\alpha$).   This change significantly impacts the shape and evolution of the afterglow light curves across all wavelengths. With $\alpha=0.0$, the afterglow exhibits a much larger flux at the beginning, sometimes exceeding two orders of magnitude compared to the panels in Figure~\ref{fig:AfterglowVar} with $\alpha=2.2$. We also note that the flux reaches its maximum faster in all bands.   This behavior indicates that energy is injected more rapidly into the system, producing higher early-time flux, especially in the X-ray band. However, in Figure \ref{fig:AfterglowVar}, the slower rise in flux leads to a more sustained injection of energy over time, resulting in a flatter decay at late times. Consequently, the light curves in Figure~\ref{fig:AfterglowVar} maintain a higher flux at later times than in Figure~\ref{fig:AfterglowVarRel}, where the steeper decline is evident due to the faster initial injection of energy.

\begin{figure}
\centering
\includegraphics[width=0.85\linewidth]{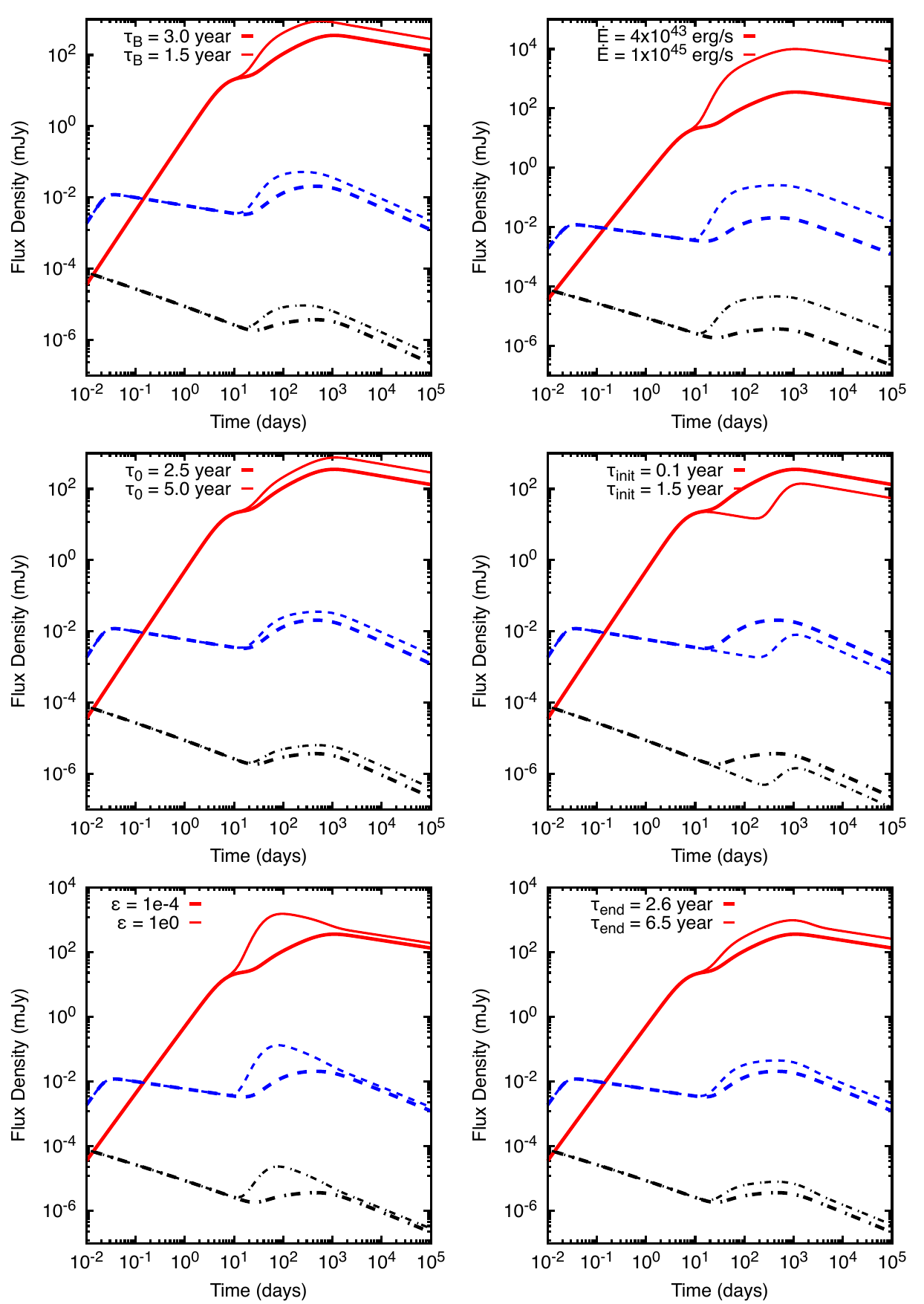}
\caption{Synchrotron light curves at X-ray (black), optical (blue) and radio (red) bands generated  by the deceleration of the non-relativistic material in the external environment. The light curves of X-ray, optical and radio bands are estimated at 1 keV, 1 eV and 10 GHz, respectively. The parameters used for the afterglow scenario are: E = $10^{49} \, {\rm erg}$, $n_{\rm s} = 1 \, {\rm cm^{-3}}$, $\varepsilon_{e} = 10^{-1}$, $\varepsilon_{B} = 10^{-2}$, $p = 2.5$, $\alpha = 2.2$, $\theta_j=4.6\,{\rm deg}$ and $\Delta \theta=15.0\,{\rm deg}$ .  Each panel progressively displays the changes of just one parameter in the emergence of B-field model. In general, these values are: $\tau_{B} = 3\, {\rm year}$, $\epsilon = 10^{-4}$, $\tau_{0} = 2.5\, {\rm year}$, $\dot{E_{0}} = 4.5\times 10^{43} \, {\rm erg/s}$. These parameter values are reported in Table \ref{tab:par_values}.}
\label{fig:MagVar_v1}
\end{figure}

\begin{figure}
\centering
\includegraphics[width=0.8\linewidth]{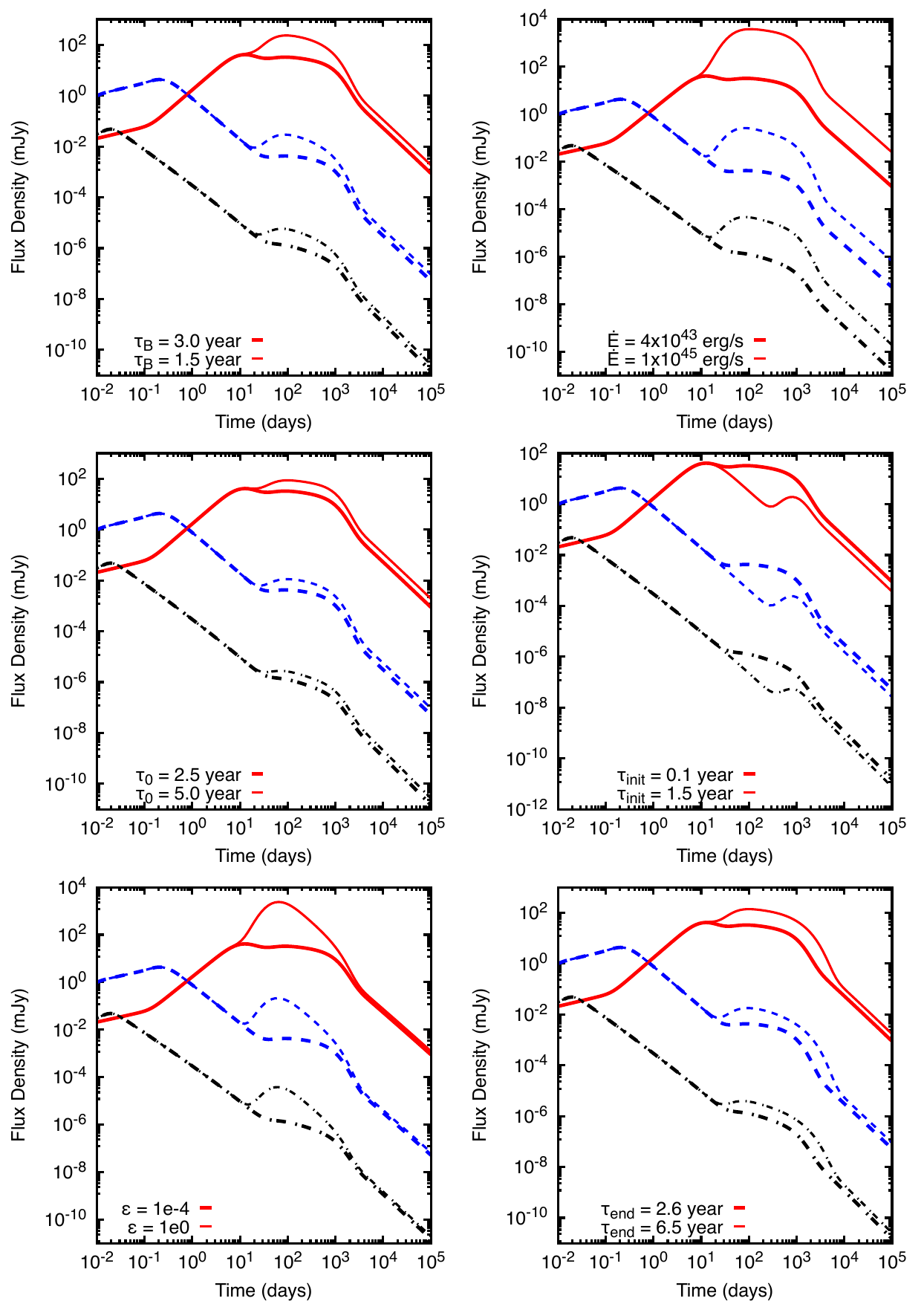}
\caption{Same as figure \ref{fig:MagVar_v1}, but for $\alpha = 0.0$.}
\label{fig:MagVarRela}
\end{figure}

In Figures~\ref{fig:MagVar_v1} and~\ref{fig:MagVarRela}, we observe the effects of changing the key parameters related to the evolution of the magnetic field, such as the diffusion timescale $\tau_B$, the initial spin-down time $\tau_0$, the magnetic field growth efficiency $\epsilon$, the initial spin-down luminosity $\dot{E}$, and the times marking the start ($\tau_{\rm init}$) and end ($\tau_{\rm end}$) of the emergence process. In the top-left panel, $\tau_B$ varies between 3.0 and 1.5 years.  A shorter $\tau_B$ leads to a faster and more pronounced rise in the late-time light curves, indicating that the magnetic field reemerges more quickly. The flux increases rapidly in all bands, but the difference is most visible in X-rays. Similarly, variations in $\tau_0$ (middle-left panel) show that a longer initial spin-down time delays the peak flux, particularly in the optical and radio bands, as the energy injection from the magnetar is sustained longer.  The panels also highlight how increasing $\epsilon$ or $\dot{E}$ (bottom-left and upper-right panels) dramatically boosts the flux, with more energetic systems producing brighter afterglows, particularly in the X-ray and optical bands. Lastly, altering $\tau_{\rm init}$ and $\tau_{\rm end}$ (middle-right and bottom-right panels) mainly shifts the time at which the peak flux is reached, with earlier or later emergence phases influencing the timing of the evolution of the synchrotron light curve. This behavior is noticeable in all bands and is most pronounced in the $\tau_{\rm init}$ panel. When we compare Figures~\ref{fig:MagVar_v1} and~\ref{fig:MagVarRela}, we see that the conclusions for figures~\ref{fig:AfterglowVar} and ~\ref{fig:AfterglowVarRel} stay unchanged.


\subsection{Discussion}

While substantial, the hypercritical accretion onto the proto magnetar does not reach the accretion rates necessary to induce gravitational collapse into a BH. Specifically, although accretion provides an additional energy source, the material deposition and resulting increase in rotational energy do not sufficiently destabilize the magnetar. Studies show that for a magnetar to collapse, the critical mass threshold would need to be exceeded under conditions that significantly increase the proto magnetar's rotational frequency or density to a level unsustainable by its internal structure \citep{rowlinson2013signatures, metzger2011protomagnetar, bernardini2013switch}. In hypercritical accretion regimes, where rates of \(\dot{M} \sim (10^{-3} - 10^{-1}) \, M_{\odot} \, \text{yr}^{-1}\) are common, the accreted mass typically does not exceed \(\sim 0.1 \, M_{\odot}\) over the relevant timescales, maintaining the system's stability \citep{chevalier1989neutron, quataert2012swift}. Thus, the rotationally supported magnetar remains below the critical point for collapse, even when accounting for the energy contributions of fallback accretion \citep{piro2011supernova, Bernal2013}. Additionally, simulations of magnetar evolution under hypercritical accretion indicate that the transient crustal submergence of the magnetic field does not alter the overall mass distribution significantly enough to disrupt the NS' equilibrium, which is crucial in delaying any potential collapse \citep{geppert1999submergence,art_vigano}. Therefore, this model suggests that hypercritical accretion onto a proto magnetar can fuel late-time activity without inevitably leading to a BH transition.\\

Since hypercritical fallback accretion onto a proto-magnetar does not inherently lead to gravitational collapse, scaling fallback parameters from newborn NSs to millisecond magnetars offers a framework for understanding delayed magnetic reactivation in these objects. The scaling process involves adjusting critical parameters such as the accretion rate, crustal composition, and thermal diffusion timescales to reflect proto-magnetars' unique energy and density profiles. For instance, in magnetars, the fallback accretion rate, typically on the order of \(\dot{M} \sim (10^{-3} - 10^{-1}) \, M_{\odot} \, \text{yr}^{-1}\) over the initial seconds to minutes post-burst, can be translated to yield comparable energy output but with extended diffusion timescales due to the greater rotational energy and magnetic field strength \citep{geppert1999submergence, art_vigano}. This modified approach implies that the crustal dynamics and field diffusion will scale with both the magnetar's initial spin rate and its intense magnetic field, allowing the reemergence timescale to be delayed by several years, consistent with the observed X-ray rebrightening in sources like GRB 170817A \citep{troja2017a, hajela2022evidence}. Importantly, our model assumes that thermomagnetic processes and crustal fractures facilitate this magnetic diffusion. These processes have been observed in simulations of NSs and can be extrapolated to the proto-magnetar case by assuming parameter values that reflect enhanced magnetic and rotational energy.  
Despite the robust fallback accretion phase, the extraordinary magnetic field strength of the magnetar introduces substantial magnetic pressures and stresses within the newly formed stellar crust. This crust, which initially formed under intense hypercritical accretion, is subject to ongoing internal stresses capable of accelerating the outward diffusion of the magnetic field. The powerful magnetic stresses within this crust increase the likelihood of crustal fractures and reconnections that enable efficient magnetic field growth and eventual reemergence over timescales of several years \citep{Bernal2013, fraija2014signatures, fraija2018hypercritical}. Thus, the interplay between the intense magnetic field and the structural pressures within the proto-magnetar supports a reemergence scenario that is physically consistent and aligned with observed reactivation phenomena in GRB afterglows.\\

We derive the dynamics of the GRB afterglow considering an electron population with a spectral index $p>2$. However, some bursts have exhibited an electron distribution with a hard spectral index in the range $1 \leq p\leq 2$ \citep[e.g., see][]{2001ApJ...558L.109D, 2024MNRAS.527.1884F}. In the previous case, our theoretical model is modified 
in the Lorentz factor of the lowest-energy electrons ($\gamma_{\rm m}$) and the respective spectral break ($\nu_{\rm m}$).  Therefore, as the magnetic field is submerged, the minimum Lorentz factor and the respective spectral break evolve as $\gamma_{\rm m}\propto t^{\frac{p(\alpha+3)-2(\alpha+5)}{2(p-1)(\alpha+3)}}$ and $\nu_{\rm m}\propto t^{-\frac{\alpha+7}{(p-1)(\alpha+3)}}$, respectively, and as the magnetic field reemerges, these evolve as $\gamma_{\rm m}\propto t^{\frac{2(\alpha-1)+n(p-2)(\alpha+3)-p(\alpha+3)}{2(n-1)(p-1)(\alpha+3)}}$ and $\nu_{\rm m}\propto t^{-\frac{5-\alpha+n(\alpha+3)}{(n-1)(p-1)(\alpha+3)}}$, respectively.\\

We consider the dynamics of the afterglow model with constant density; however, in some cases, the best-fit afterglow model at the timescale of months has favored a stellar wind instead of a homogeneous medium \citep[e.g., see][]{2023MNRAS.525.1630F, 2022ApJ...934..188F}.   In the afterglow wind scenario, the density profile can be modeled by $n(r) =A r^{\rm -2}$ with $A=A_{\rm \star}\frac{\dot{M}_{\rm W}}{4\pi m_p v_{\rm W}}=A_{\rm \star}3.0\times 10^{35}\,{\rm cm^{-1}}$, where $\dot{M}_{\rm W}$,  $v_{\rm W}$ and $m_p$ are the mass-loss rate, the wind velocity and the proton mass, respectively \citep{2000ApJ...536..195C, 1998ApJ...501..772P}.  As the magnetic field is submerged by hypercritical accretion, the minimum and cooling Lorentz factors evolve as $\gamma_{\rm m}\propto t^{-\frac{2}{\alpha+3}}$ and $\gamma_{\rm c}\propto t$, respectively. The respective synchrotron spectral breaks vary as $\nu_{\rm m}\propto t^{-\frac{\alpha+7}{\alpha+3}}$,  $\nu_{\rm c}\propto t$, and the maximum synchrotron flux as $F_{\rm max}\propto t^{-\frac{1}{\alpha+3}}$. During the influence of the magnetic field's growth, the minimum and cooling Lorentz factor evolve as $\gamma_{\rm m}\propto t^{-\frac{4}{(n-1)(\alpha+3)}}$ and $\gamma_{\rm c}\propto t$, respectively. The respective synchrotron spectral breaks vary as $\nu_{\rm m}\propto t^{-\frac{5+3n-\alpha(1-n)}{(n-1)(\alpha+3)}}$,  $\nu_{\rm c}\propto t$, and the maximum synchrotron flux as $F_{\rm max}\propto t^{-\frac{2}{(n-1)(\alpha+3)}}$.\\

We derive in Tables \ref{Table2} and \ref{Table3}, the closure relations that describe the evolution of synchrotron flux ($F_\nu \propto t^{-\alpha_c}\nu^{-\beta_c}$) with and without the influence of the magnetic field’s growth.  The closure relations are the equations that determine the connections between the spectral PL index ($\beta_c$) and the temporal PL index ($\alpha_c$) of a specific segment of the lightcurves. According to the flux evolution, $F_v \propto t^{-\alpha_c}v^{-\beta}$, closure relations could help us to infer the afterglow physics as the circumburst density (wind vs homogeneous environment), hydrodynamics  (adiabatic vs radiative), energy injection (continuous vs instantaneous), variation of microphysical parameters among others. Numerous studies have been performed across various wavelengths in high-energy gamma rays, X-rays, and optical spectra. 
 For instance, in the gamma-ray energies, 
\cite{2019ApJ...883..134T,2021ApJS..255...13D,Dainotti2023Galax..11...25D,2023ApJ...958..126F,2024MNRAS.527.1884F,2024MNRAS.527.1674F} investigated the closure relations considering as a GRB sample those reported in the Second Fermi-LAT GRB Catalog \citep[2FLGC][]{2019ApJ...878...52A}. In X-ray bands, \citet{2009ApJ...698...43R}, \citet{Srinivasaragavan2020ApJ}, and \citet{2021PASJ...73..970D} investigated the closure relations considering most of the bursts reported in the Swift-XRT database. In optical bands, \citet{oates2012} considered 48 bursts observed by the Swift-UVOT instrument. Most of these studies revealed that a scenery with energy injection scenario was more favorable than one without energy injection, all lying in the slow-cooling regime. Although different parameter values of our scenario might lead to much smaller timescales of submergence and the emergence of magnetic fields,  this analysis is beyond the scope of the current paper.\\

Due to insufficient understanding of the energy transfer mechanisms between protons, electrons, and magnetic fields in relativistic shocks, evolution of the magnetic field during the GRB afterglows has been considered through variations in the values of the microphysical parameter ($\varepsilon_{\rm B}\propto t^{-b}$) \citep[e.g., see][]{2003ApJ...597..459Y, 2003MNRAS.346..905K, 2006A&A...458....7I,2006MNRAS.369..197F, 2006MNRAS.369.2059P, 2005PThPh.114.1317I, 2006MNRAS.370.1946G, 2013MNRAS.428..845L, 2020ApJ...905..112F}. \cite{2013MNRAS.428..845L} considered that the magnetic field could be amplified due to the quick and long evolution of microturbulences. The authors modeled the multiwavelength observation in some bursts detected by Fermi-LAT. \cite{2006MNRAS.370.1946G} proposed that the plateau phase may be elucidated by the variation of the magnetic microphysical parameter instead of the energy injection. To interpret the prominent peaks exhibited in several GRB afterglows, \cite{10.1111/j.1365-2966.2009.15886.x} considered the evolution of these microphysical parameters in the scenario of a wind bubble. \cite{2020ApJ...905..112F} proposed that the breaks observed in different bands (e.g. GeV gamma rays and X-rays) could be correlated through the rapid evolution of this microphysical parameter.  Using the microphysical parameter variations of the synchrotron afterglow model, recently \cite{2024MNRAS.527.1884F} calculated the closure relations and light curves in a stratified medium.    We propose that the evolution of the magnetic field could be explained via the hypercritical accretion, and the effect of growth due to the reemergence in a shorter timescale.\\

\cite{2020MNRAS.492.5011D} conducted a systematic search for nearby short gamma-ray bursts (sGRBs) that exhibited characteristics equivalent to GRB 170817A within the Swift database, covering a 14-year operational period from 2005 to 2019, as documented in the online BAT GRB catalog \citep{2016ApJ...829....7L}. A subset of four potential bursts (GRB 050906, 070810B, 080121, and 100216A) was reported in a range of $100\leq d_z\leq 200\,{\rm Mpc}$.  The range of parameters for the X-ray counterparts of two NSs merging was constrained using these bursts, and optical upper limits in a timescale of days were derived. These bursts with similar features of GRB 170817A represent potential candidates so that the hypercritical accretion could have submerged the magnetic field. A similar conclusion may be stated around the GW event GW230529\_181500, which was reported on 2023 May 29, during the fourth observation run (O4) \citep{2024ApJ...970L..34A}. Nonelectromagnetic signatures were associated with this GW event, although a compact binary coalescence with masses estimated in ranges of 2.5 - 4.5 $M_\odot$ and 1.2 and 2.0 $M_\odot$ was predicted.\\

Several GRBs exhibited characteristics that could be explained by the scenario of a millisecond magnetar undergoing hypercritical accretion, followed by the submergence and reemergence of its magnetic field. For instance, GRB 200415A, located at the nearby galaxy NGC 253, showcased a rapid onset and fast variability, which could indicate a magnetar flare. The properties of this GRB, including its energetic and spectral characteristics, align with predictions for giant flares from magnetars experiencing dynamic magnetic field interactions \citep{Roberts2021Rapid}. Another plausible candidate is GRB 080503, which presented an interesting case where the scenario of a millisecond magnetar with hypercritical accretion could be applicable, particularly in explaining its late afterglow and rebrightening features. Recent research indicates that GRB 080503's properties, including a late optical and X-ray rebrightening, align well with predictions from millisecond magnetar models, especially when considering magnetar-powered "merger-nova" emissions. These phenomena suggest that the remnant of a double NS merger could be a rapidly rotating, highly magnetized NS (millisecond magnetar), which significantly influences the observed afterglow characteristics through its energy output and magnetic dynamics \citep{Gao2015GRB}.  \cite{ber13} compiled a collection of eight gamma-ray bursts (GRBs) with established redshifts identified by Swift/BAT from 2005 to 2009. This collection not only exhibited indications of late activity characterized by a plateau but also demonstrated evidence of precursor activity. This condition is explicable solely through accretion processes if the central engine is identified as a newly formed magnetar.   \cite{lu14} presented a sample of 214 candidates for magnetar central engines with established redshifts, categorizing them into four groups (Gold, Silver, Aluminum, and others) according to the likelihood of a magnetar generating one of these bursts.   \cite{li18} applied analogous reasoning to subclassify candidates derived from the \textit{Swift/XRT} light curves. The sample comprises 101 progenitors categorized into Gold, Silver, and Bronze based on an external plateau.\\

The current model could be extended to soft gamma-ray repeaters, which share similarities with classical GRBs. For instance, SGR 1806-20, a giant flare from this soft gamma-ray repeater, exhibited an intense initial pulse followed by a fading tail. This can be understood through the framework of a temporarily submerged magnetic field that emerges to release a significant amount of energy \citep{2005Natur.434.1098H}.

\section{Particular case: GRB 170817A}\label{sec4:model}

The gravitational wave (GW) event known as GW170817 \citep{PhysRevLett.119.161101,2041-8205-848-2-L12}, which occurred on August 17, 2017, was associated with the weak gamma-ray prompt emission of GRB 170817A \citep{2017ApJ...848L..14G, 2017ApJ...848L..15S}.   The gravitational wave signal originating from the elliptical galaxy NGC 4993 at a distance of $\sim 40 {\rm Mpc}$ aligned perfectly with the predictions for the merger of two NSs, occurring almost two seconds prior to GRB 170817A. An extended observing effort spanning radio, optical, and X-ray wavelengths followed GRB 170817A \citep[e.g., see][and references therein]{troja2017a, 2041-8205-848-2-L12, 2017arXiv171100243K, 2018arXiv180106164D}. Numerous authors modeled the observational data of the nonthermal spectrum of GRB 170817A collected during the initial 900 days following the merger event. It was demonstrated that these data were congruent with synchrotron forward-shock emission produced by the slowing of an off-axis structured jet viewed at $15^{\circ}\leq \theta_{\mathrm{obs}}\leq 25^{\circ}$ \citep{troja2017a, 2017Sci...358.1559K, 2017MNRAS.472.4953L,  2017ApJ...848L..20M, 2017ApJ...848L...6L, 2018MNRAS.479..588G, 2019ApJ...884...71F, 2018MNRAS.479..588G, 2018ApJ...867...95H, 2019ApJ...871..200F}. Some authors argued that the off-axis structured jet originated from an off-axis jet and quasi-spherical materials \citep{2017ApJ...848L...6L, 2018MNRAS.479..588G, 2019ApJ...884...71F, 2018MNRAS.479..588G, 2018ApJ...867...95H, 2019ApJ...871..200F,2021MNRAS.503.4363U}.  The observational data were detected in the following 3.3 years with the Chandra X-ray Observatory, the Very Large Array (VLA), and the MeerKAT radio interferometer. \cite{2022ApJ...927L..17H} examined these observations and determined that the data were inconsistent with the appropriate synchrotron curves from the off-axis jet model, providing evidence of an unexpected X-ray excess. In light of these divergent data, the authors proposed an approach to elucidate this occurrence within either radiation from accretion processes on the compact object remnant or a decelerated non-relativistic quasi-spherical material.

\begin{figure}
\centering
\includegraphics[width=\linewidth]{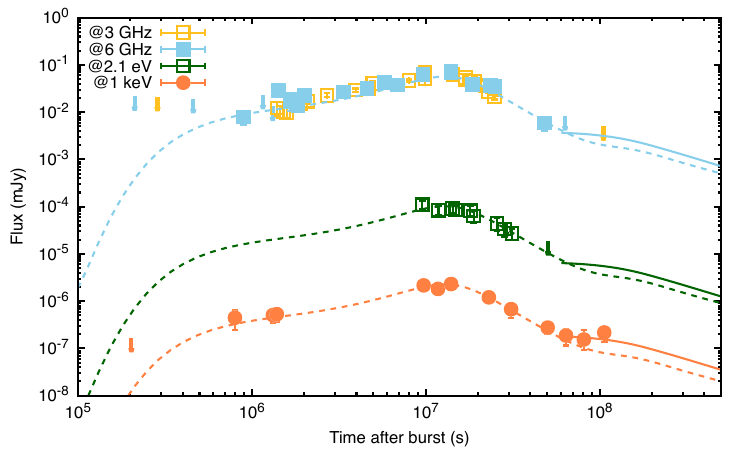}
\caption{Multiwavelength observations of GRB 170817A with the synchrotron light curves from decelerated materials  with and without the emergence of magnetic field.  The radio wavelengths at 3 and 6 GHz are shown in yellow and blue, respectively, the optical band exhibited at 2.1 eV is in green, and the X-rays at 1 keV is in orange.   The dashed lines correspond to the best-fit curves using MCMC simulations from the analytical model of the synchrotron emission for an off-axis top-hat jet and a quasi-spherical outflow reported in \citep{2019ApJ...884...71F}, and the solid lines are obtained with a test statistic $\chi^{2} = 1.14$ taking into account the emergence of the magnetic field. The best-fit parameter values are reported in Table \ref{tab:par_values}.}
\label{fig:GRB170817A}
\end{figure}

Figure \ref{fig:GRB170817A} presents the radio, optical, and X-ray observations and upper limits of GRB 170817A, together with the best-fit synchrotron lightcurves. The best-fit curves (solid lines) of the magnetic field reemergence model described in Section~\ref{sec3:model} were performed with the \texttt{LMFIT} \citep{newville_matthew_2014_11813} Python package, taking into account the X-ray data at 1 keV after 500 days since the burst, allowing the parameters $\tau_{0}$, $\tau_{\rm B}$, $\epsilon$ and $\dot(E)_{0}$ to be free.  The method to minimize the model was 'nelder'. Once the fit was completed, we extrapolated the results to the optical and radio bands.   The dashed lines correspond to the best-fit curves from the afterglow model of the synchrotron emission for an off-axis top-hat jet and a quasi-spherical outflow reported in \cite{2019ApJ...884...71F}. The fit parameters of the afterglow model ($E$, $n$, $p$, $\theta_j$, $\Delta \theta$, $\varepsilon_B$, $\varepsilon_e$ and $\alpha_s$) was performed with a Markov chain Monte Carlo (MCMC) code \citep[see,][]{2019ApJ...871..200F}. A total of $17600$ samples were performed with $5150$ tuning steps. The best-fit values of the emergence of the magnetic field scenario and afterglow model are reported in Table \ref{tab:par_values}. It is worth noting that the equations with the set of parameters are degenerate, meaning that identical findings could be obtained for an entirely other set of values. As a result, our finding is only one potential solution and is not unique. This fit implies a more intense and shorter duration of the late-time rebrightening, which may reflect a rapid reconfiguration of the magnetic field driving a faster spin-down.

\subsection{A long-lived NS as a remnant of GW170817} 

Three years after the merger of an NS binary, X-ray observations began to show an excess over the forward-shock jet models' predictions. As a result, models with an extra X-ray source started to appear. Although these models are only preferred at $1\sigma-2\sigma$, \cite{2024ApJ...975..131R} note that they better fit the data than their all-inclusive single-source afterglow model. More observations of this occurrence are required to determine if a new component is present. \cite{2024ApJ...971L..24D} proposed a model that implied the existence of a relativistic equatorial outflow generated by a long-lived NS remnant. This outflow may carry a sizable portion of the remnant's rotational energy. Due to relativistic beaming, this emission would first be obscured from view, but when the outflow slows down, and its beaming cone widens, it may become visible and appear as the X-ray excess.   Given that the remnant of the binary NSs of GRB 170817A was a revolving, highly magnetized NS, \cite{2021ApJ...918...52L} demonstrated that late multiwavelength data may be explained by a relativistic electron/positron pair wind with an off-axis top-hat jet.
Furthermore, \cite{2021MNRAS.506.5908N} showed that high-mass ratio models with stiff equations of state and equal mass models with soft equations of state are less preferred because they typically predict afterglows that peak too quickly to explain the most recent data. Generally, the models with moderate mass ratios and stiffness agreed with the most recent findings. The evolution of multiwavelength radiation from a KN ejecta-pulsar wind nebula (PWN) system was examined by \cite{10.1093/mnras/stac797}. The authors demonstrated how a PWN driven by the NS remnant following GW170817 might re-brighten the late-time X-ray afterglow of GRB 170817A and alter the optical transient AT 2017gfo.   Consistent with our magnetic field scenario, these models indicate that the GW170817 event left behind a long-lived NS.


\section{Conclusions}\label{sec5:model}


Millisecond magnetars remain a compelling candidate for the central engine of gamma-ray bursts (GRBs), influencing both the prompt energetics and the long-term afterglow \citep[e.g.][]{Beniamini2017Constraints}. Their magnetic-dipole spin-down can inject power over $\sim 10^3$--$10^5$\,s, producing the well-known X-ray plateaus, while hypercritical fallback may submerge the external field and delay that energy release \citep{Sarin2020Interpreting,Bucciantini2011Magnetars}. 

In this work, we have developed an analytic framework that couples magnetic submergence, Hall--Ohm diffusion, and crustal reconnection to a non-relativistic afterglow model, and we have applied it to the late-time brightening of GW170817/GRB~170817A.

\smallskip
\noindent\textbf{Key quantitative conclusions.}
\begin{enumerate}
\item \textbf{Diffusion timescale:}  
Joint X-ray and radio fitting constrains the re-emergence timescale to $3 \lesssim \tau_{B} \lesssim 40$\,yr. The lower bound corresponds to the high-field, Hall-dominated limit ($B \gtrsim 3 \times 10^{15}$\,G), while the upper bound reflects more typical magnetar fields in which Ohmic diffusion dominates.

\item \textbf{Burial depth:}  
The plateau amplitude implies that at least 90\,\% of the external dipole flux was submerged, i.e.\ $\epsilon = (B_0 / B_{\max})^2 \simeq 10^{-3}$--$10^{-2}$, consistent with an accreted mass of $\sim 10^{-2}\,M_{\odot}$.

\item \textbf{Recovered field strength:}  
Once diffusion saturates, the surface dipole is restored to $B_{\mathrm{surf}} \simeq (2$--$5) \times 10^{15}$\,G, consistent with the magnetar-level fields inferred from Galactic transients.

\item \textbf{Strength of evidence:}  
When burial is included, the model reproduces both the late-time rise in the X-ray flux and the radio spectral slope with a single additional parameter ($\tau_{B}$). In contrast, a no-burial model underpredicts the X-ray flux at 1000--2000\,d by more than an order of magnitude, even after re-optimising all other parameters. The re-emergence scenario therefore provides the most economical explanation for the decade-long brightening observed in GRB~170817A.

\end{enumerate}

\paragraph*{Alternative plateau models.}
Inclination-angle evolution or free precession \citep{SuvorovKokkotas2020} can produce a quasi-flat plateau, but such engines fade once the wobble damps. In contrast, the diffusion model predicts a late-time brightening as the dipole resurfaces. The observed rise in X-ray flux beyond 1000\,d therefore favours magnetic re-emergence, unless the damping timescale of precession were fine-tuned to several years. Detection of periodic X-ray modulations or continuous gravitational waves would provide a decisive test between the two scenarios.

\paragraph*{Robustness to stellar structure.}
Varying the NS radius within 11--14.5\,km (corresponding to $I = (1.1$--$2.0)\times 10^{45}$\,g\,cm$^2$; \citealt{Riley2021}) simply rescales the luminosity normalization. Since the diffusion parameters $\tau_B$ and $\epsilon$ are tied to the timing and shape of the late-time bump, they remain uniquely determined: changing $I$ by 30\,\% shifts $\dot{E}_0$ by the same factor, but alters $\tau_B$ by less than 10\,\%. 
This analytic argument, already noted by
\citet{DallOsso2011} and \citet{Rowlinson2013}, shows that degeneracy
with $R_\star$ and $I$ does not compromise the main results, so we
have not added extra structural parameters to the numerical fits. We therefore conclude that the evidence for magnetic re-emergence is robust against plausible uncertainties in NS structure.

\medskip
\noindent
In summary, magnetic-field submergence and re-emergence are key components of the millisecond-magnetar paradigm, capable of explaining both plateau phases and late-time brightening in short and long GRBs.
Extending this analysis to a larger sample and performing multi-dimensional MHD simulations will be essential to refine the diffusion timescale, assess the role of plastic flow, and further test the magnetar reactivation hypothesis.


\section*{Acknowledgements}

We express our gratitude to the anonymous referee for their meticulous assessment of the paper and insightful recommendations, which greatly enhanced the quality and clarity of our paper. The authors thank Tanmoy Laskar and Peter Veres for useful discussions. NF acknowledges financial support from UNAM-DGAPA-PAPIIT through the grant IN112525.  AG is grateful to UNAM-DGAPA-PAPIIT. This work was supported by Universidad Nacional Autónoma de México Postdoctoral Program (POSDOC). The software used in this work was developed in part by the DOE NNSA- and DOE Office of Science-supported Flash Center for Computational Science at the University of Chicago and the University of Rochester.

\section*{Data Availability}


There are no new data associated with this article.



\bibliographystyle{mnras}
\bibliography{Emergence_Bfield_v3} 

\begin{thebibliography}{}
\makeatletter
\relax
\def\mn@urlcharsother{\let\do\@makeother \do\$\do\&\do\#\do\^\do\_\do\%\do\~}
\def\mn@doi{\begingroup\mn@urlcharsother \@ifnextchar [ {\mn@doi@}
  {\mn@doi@[]}}
\def\mn@doi@[#1]#2{\def\@tempa{#1}\ifx\@tempa\@empty \href
  {http://dx.doi.org/#2} {doi:#2}\else \href {http://dx.doi.org/#2} {#1}\fi
  \endgroup}
\def\mn@eprint#1#2{\mn@eprint@#1:#2::\@nil}
\def\mn@eprint@arXiv#1{\href {http://arxiv.org/abs/#1} {{\tt arXiv:#1}}}
\def\mn@eprint@dblp#1{\href {http://dblp.uni-trier.de/rec/bibtex/#1.xml}
  {dblp:#1}}
\def\mn@eprint@#1:#2:#3:#4\@nil{\def\@tempa {#1}\def\@tempb {#2}\def\@tempc
  {#3}\ifx \@tempc \@empty \let \@tempc \@tempb \let \@tempb \@tempa \fi \ifx
  \@tempb \@empty \def\@tempb {arXiv}\fi \@ifundefined
  {mn@eprint@\@tempb}{\@tempb:\@tempc}{\expandafter \expandafter \csname
  mn@eprint@\@tempb\endcsname \expandafter{\@tempc}}}

\bibitem[\protect\citeauthoryear{{Abac} et~al.,}{{Abac}
  et~al.}{2024}]{2024ApJ...970L..34A}
{Abac} A.~G.,  et~al., 2024, \mn@doi [\apjl] {10.3847/2041-8213/ad5beb}, \href
  {https://ui.adsabs.harvard.edu/abs/2024ApJ...970L..34A} {970, L34}

\bibitem[\protect\citeauthoryear{Abbott, Abbott, Abbott  \& et al.}{Abbott
  et~al.}{2017a}]{PhysRevLett.119.161101}
Abbott B.~P.,  Abbott R.,  Abbott T.~D.,   et al. 2017a, \mn@doi [Phys. Rev.
  Lett.] {10.1103/PhysRevLett.119.161101}, 119, 161101

\bibitem[\protect\citeauthoryear{Abbott, Abbott, Abbott  \& et al.}{Abbott
  et~al.}{2017b}]{2041-8205-848-2-L12}
Abbott B.~P.,  Abbott R.,  Abbott T.~D.,   et al. 2017b, The Astrophysical
  Journal Letters, 848, L12

\bibitem[\protect\citeauthoryear{Aguilera, Pons  \& Miralles}{Aguilera
  et~al.}{2008}]{Aguilera2008}
Aguilera D.~N.,  Pons J.~A.,   Miralles J.~A.,  2008, Astronomy \&
  Astrophysics, 486, 255

\bibitem[\protect\citeauthoryear{{Ajello} et~al.,}{{Ajello}
  et~al.}{2019}]{2019ApJ...878...52A}
{Ajello} M.,  et~al., 2019, \mn@doi [\apj] {10.3847/1538-4357/ab1d4e}, \href
  {https://ui.adsabs.harvard.edu/abs/2019ApJ...878...52A} {878, 52}

\bibitem[\protect\citeauthoryear{{Barkov} \& {Komissarov}}{{Barkov} \&
  {Komissarov}}{2008}]{2008MNRAS.385L..28B}
{Barkov} M.~V.,  {Komissarov} S.~S.,  2008, \mn@doi [\mnras]
  {10.1111/j.1745-3933.2008.00427.x}, \href
  {https://ui.adsabs.harvard.edu/abs/2008MNRAS.385L..28B} {385, L28}

\bibitem[\protect\citeauthoryear{Beniamini, Giannios  \& Metzger}{Beniamini
  et~al.}{2017}]{Beniamini2017Constraints}
Beniamini P.,  Giannios D.,   Metzger B.,  2017, \mn@doi [Monthly Notices of
  the Royal Astronomical Society] {10.1093/mnras/stx2095}, 472, 3058

\bibitem[\protect\citeauthoryear{Bernal, Lee  \& Page}{Bernal
  et~al.}{2010}]{bernal2010hypercritical}
Bernal C.~G.,  Lee W.~H.,   Page D.,  2010, Hypercritical accretion onto a
  magnetized neutron star surface: a numerical approach (\mn@eprint {arXiv}
  {1006.3003})

\bibitem[\protect\citeauthoryear{{Bernal}, {Page}  \& {Lee}}{{Bernal}
  et~al.}{2013a}]{2013ApJ...770..106B}
{Bernal} C.~G.,  {Page} D.,   {Lee} W.~H.,  2013a, The Astrophysical Journal,
  770, 106

\bibitem[\protect\citeauthoryear{Bernal, Page  \& Lee}{Bernal
  et~al.}{2013b}]{Bernal2013}
Bernal C.~G.,  Page D.,   Lee W.~H.,  2013b, The Astrophysical Journal, 770,
  106

\bibitem[\protect\citeauthoryear{Bernal, Frajuca, Hirsch, Minari, Magalhaes  \&
  Selbach}{Bernal et~al.}{2024}]{bernal2024overall}
Bernal C.~G.,  Frajuca C.,  Hirsch H.~D.,  Minari B.,  Magalhaes N.~S.,
  Selbach L.~B.,  2024, Frontiers in Astronomy and Space Sciences, 11, 1390597

\bibitem[\protect\citeauthoryear{Bernardini et~al.,}{Bernardini
  et~al.}{2013a}]{bernardini2013switch}
Bernardini M.~G.,  et~al., 2013a, The Astrophysical Journal, 775, 67

\bibitem[\protect\citeauthoryear{Bernardini et~al.,}{Bernardini
  et~al.}{2013b}]{ber13}
Bernardini M.,  et~al., 2013b, The Astrophysical Journal, 775, 67

\bibitem[\protect\citeauthoryear{Blandford \& Romani}{Blandford \&
  Romani}{1988}]{art_blandford}
Blandford R.~D.,  Romani R.~W.,  1988, Monthly Notices of the Royal
  Astronomical Society, 234, 57P

\bibitem[\protect\citeauthoryear{{Blandford} \& {Znajek}}{{Blandford} \&
  {Znajek}}{1977}]{1977MNRAS.179..433B}
{Blandford} R.~D.,  {Znajek} R.~L.,  1977, \mn@doi [\mnras]
  {10.1093/mnras/179.3.433}, \href
  {http://adsabs.harvard.edu/abs/1977MNRAS.179..433B} {179, 433}

\bibitem[\protect\citeauthoryear{Bucciantini}{Bucciantini}{2011}]{Bucciantini2011Magnetars}
Bucciantini N.,  2011, \mn@doi [Proceedings of the International Astronomical
  Union] {10.1017/S1743921312013075}, 7, 289

\bibitem[\protect\citeauthoryear{{Cannizzo} \& {Gehrels}}{{Cannizzo} \&
  {Gehrels}}{2009}]{Cannizzo2009}
{Cannizzo} J.~K.,  {Gehrels} N.,  2009, \mn@doi [\apj]
  {10.1088/0004-637X/700/2/1047}, \href
  {http://adsabs.harvard.edu/abs/2009ApJ...700.1047C} {700, 1047}

\bibitem[\protect\citeauthoryear{{Cannizzo}, {Troja}  \& {Gehrels}}{{Cannizzo}
  et~al.}{2011}]{cannizzo2011}
{Cannizzo} J.~K.,  {Troja} E.,   {Gehrels} N.,  2011, \mn@doi [\apj]
  {10.1088/0004-637X/734/1/35}, \href
  {http://adsabs.harvard.edu/abs/2011ApJ...734...35C} {734, 35}

\bibitem[\protect\citeauthoryear{Chamel \& Haensel}{Chamel \&
  Haensel}{2008}]{ChamelHaensel2008}
Chamel N.,  Haensel P.,  2008, Living Rev. Relativ., 11, 10

\bibitem[\protect\citeauthoryear{{Chevalier}}{{Chevalier}}{1989a}]{1989ApJ...346..847C}
{Chevalier} R.~A.,  1989a, \mn@doi [\apj] {10.1086/168066}, \href
  {https://ui.adsabs.harvard.edu/abs/1989ApJ...346..847C} {346, 847}

\bibitem[\protect\citeauthoryear{Chevalier}{Chevalier}{1989b}]{chevalier1989neutron}
Chevalier R.~A.,  1989b, Astrophysical Journal, 346, 847

\bibitem[\protect\citeauthoryear{Chevalier \& Emmering}{Chevalier \&
  Emmering}{1986}]{chevalier1986pulsars}
Chevalier R.~A.,  Emmering R.~T.,  1986, Astrophysical Journal, 304, 140

\bibitem[\protect\citeauthoryear{{Chevalier} \& {Li}}{{Chevalier} \&
  {Li}}{2000}]{2000ApJ...536..195C}
{Chevalier} R.~A.,  {Li} Z.-Y.,  2000, \mn@doi [\apj] {10.1086/308914}, \href
  {http://adsabs.harvard.edu/abs/2000ApJ...536..195C} {536, 195}

\bibitem[\protect\citeauthoryear{{Choudhuri} \& {Konar}}{{Choudhuri} \&
  {Konar}}{2002}]{2002MNRAS.332..933C}
{Choudhuri} A.~R.,  {Konar} S.,  2002, \mn@doi [\mnras]
  {10.1046/j.1365-8711.2002.05362.x}, \href
  {https://ui.adsabs.harvard.edu/abs/2002MNRAS.332..933C} {332, 933}

\bibitem[\protect\citeauthoryear{Chugunov \& Horowitz}{Chugunov \&
  Horowitz}{2010}]{chugunov2010breaking}
Chugunov A.~I.,  Horowitz C.~J.,  2010, Monthly Notices of the Royal
  Astronomical Society: Letters, 407, L54

\bibitem[\protect\citeauthoryear{Cumming, Arras  \& Zweibel}{Cumming
  et~al.}{2004}]{cumming2004magnetic}
Cumming A.,  Arras P.,   Zweibel E.,  2004, The Astrophysical Journal, 609, 999

\bibitem[\protect\citeauthoryear{Cumming, Macbeth, Page  et~al.}{Cumming
  et~al.}{2006}]{cumming2006long}
Cumming A.,  Macbeth J.,  Page D.,   et~al., 2006, The Astrophysical Journal,
  646, 429

\bibitem[\protect\citeauthoryear{{D'Avanzo} et~al.,}{{D'Avanzo}
  et~al.}{2018}]{2018arXiv180106164D}
{D'Avanzo} P.,  et~al., 2018, preprint, \href
  {http://adsabs.harvard.edu/abs/2018arXiv180106164D} {} (\mn@eprint {arXiv}
  {1801.06164})

\bibitem[\protect\citeauthoryear{{Dai} \& {Cheng}}{{Dai} \&
  {Cheng}}{2001}]{2001ApJ...558L.109D}
{Dai} Z.~G.,  {Cheng} K.~S.,  2001, \mn@doi [\apjl] {10.1086/323566}, \href
  {https://ui.adsabs.harvard.edu/abs/2001ApJ...558L.109D} {558, L109}

\bibitem[\protect\citeauthoryear{{Dainotti}, {Lenart}, {Fraija}, {Nagataki},
  {Warren}, {De Simone}, {Srinivasaragavan}  \& {Mata}}{{Dainotti}
  et~al.}{2021a}]{2021PASJ...73..970D}
{Dainotti} M.~G.,  {Lenart} A.~{\L}.,  {Fraija} N.,  {Nagataki} S.,  {Warren}
  D.~C.,  {De Simone} B.,  {Srinivasaragavan} G.,   {Mata} A.,  2021a, \mn@doi
  [\pasj] {10.1093/pasj/psab057}, \href
  {https://ui.adsabs.harvard.edu/abs/2021PASJ...73..970D} {73, 970}

\bibitem[\protect\citeauthoryear{{Dainotti} et~al.,}{{Dainotti}
  et~al.}{2021b}]{2021ApJS..255...13D}
{Dainotti} M.~G.,  et~al., 2021b, \mn@doi [\apjs] {10.3847/1538-4365/abfe17},
  \href {https://ui.adsabs.harvard.edu/abs/2021ApJS..255...13D} {255, 13}

\bibitem[\protect\citeauthoryear{{Dainotti}, {Levine}, {Fraija}, {Warren},
  {Veres}  \& {Sourav}}{{Dainotti} et~al.}{2023}]{Dainotti2023Galax..11...25D}
{Dainotti} M.,  {Levine} D.,  {Fraija} N.,  {Warren} D.,  {Veres} P.,
  {Sourav} S.,  2023, \mn@doi [Galaxies] {10.3390/galaxies11010025}, \href
  {https://ui.adsabs.harvard.edu/abs/2023Galax..11...25D} {11, 25}

\bibitem[\protect\citeauthoryear{Dall'Osso, Granot  \& Piran}{Dall'Osso
  et~al.}{2011}]{DallOsso2011}
Dall'Osso S.,  Granot J.,   Piran T.,  2011, Mon. Not. R. Astron. Soc., 415,
  191

\bibitem[\protect\citeauthoryear{{Dichiara}, {Troja}, {O'Connor}, {Marshall},
  {Beniamini}, {Cannizzo}, {Lien}  \& {Sakamoto}}{{Dichiara}
  et~al.}{2020}]{2020MNRAS.492.5011D}
{Dichiara} S.,  {Troja} E.,  {O'Connor} B.,  {Marshall} F.~E.,  {Beniamini} P.,
   {Cannizzo} J.~K.,  {Lien} A.~Y.,   {Sakamoto} T.,  2020, \mn@doi [\mnras]
  {10.1093/mnras/staa124}, \href
  {https://ui.adsabs.harvard.edu/abs/2020MNRAS.492.5011D} {492, 5011}

\bibitem[\protect\citeauthoryear{{DuPont} \& {MacFadyen}}{{DuPont} \&
  {MacFadyen}}{2024}]{2024ApJ...971L..24D}
{DuPont} M.,  {MacFadyen} A.,  2024, \mn@doi [\apjl]
  {10.3847/2041-8213/ad66d2}, \href
  {https://ui.adsabs.harvard.edu/abs/2024ApJ...971L..24D} {971, L24}

\bibitem[\protect\citeauthoryear{{Duncan} \& {Thompson}}{{Duncan} \&
  {Thompson}}{1992}]{1992ApJ...392L...9D}
{Duncan} R.~C.,  {Thompson} C.,  1992, \mn@doi [\apjl] {10.1086/186413}, \href
  {https://ui.adsabs.harvard.edu/abs/1992ApJ...392L...9D} {392, L9}

\bibitem[\protect\citeauthoryear{{Fan} \& {Piran}}{{Fan} \&
  {Piran}}{2006}]{2006MNRAS.369..197F}
{Fan} Y.,  {Piran} T.,  2006, \mnras, \href
  {https://ui.adsabs.harvard.edu/abs/2006MNRAS.369..197F} {369, 197}

\bibitem[\protect\citeauthoryear{Fraija \& Bernal}{Fraija \&
  Bernal}{2015}]{fraija2015hypercritical}
Fraija N.,  Bernal C.~G.,  2015, Monthly Notices of the Royal Astronomical
  Society, 451, 455

\bibitem[\protect\citeauthoryear{Fraija, Bernal  \& Hidalgo-Gamez}{Fraija
  et~al.}{2014}]{fraija2014signatures}
Fraija N.,  Bernal C.~G.,   Hidalgo-Gamez A.~M.,  2014, Monthly Notices of the
  Royal Astronomical Society, 442, 239

\bibitem[\protect\citeauthoryear{Fraija, Bernal, Morales  \& Negreiros}{Fraija
  et~al.}{2018}]{fraija2018hypercritical}
Fraija N.,  Bernal C.,  Morales G.,   Negreiros R.,  2018, Physical Review D,
  98, 083012

\bibitem[\protect\citeauthoryear{{Fraija}, {Pedreira}  \& {Veres}}{{Fraija}
  et~al.}{2019a}]{2019ApJ...871..200F}
{Fraija} N.,  {Pedreira} A.~C.~C.~d.~E.~S.,   {Veres} P.,  2019a, \mn@doi
  [\apj] {10.3847/1538-4357/aaf80e}, \href
  {http://adsabs.harvard.edu/abs/2019ApJ...871..200F} {871, 200}

\bibitem[\protect\citeauthoryear{{Fraija}, {Lopez-Camara}, {Pedreira},
  {Betancourt Kamenetskaia}, {Veres}  \& {Dichiara}}{{Fraija}
  et~al.}{2019b}]{2019ApJ...884...71F}
{Fraija} N.,  {Lopez-Camara} D.,  {Pedreira} A.~C. C. d. E.~S.,  {Betancourt
  Kamenetskaia} B.,  {Veres} P.,   {Dichiara} S.,  2019b, \mn@doi [\apj]
  {10.3847/1538-4357/ab40a9}, \href
  {https://ui.adsabs.harvard.edu/abs/2019ApJ...884...71F} {884, 71}

\bibitem[\protect\citeauthoryear{{Fraija}, {Laskar}, {Dichiara}, {Beniamini},
  {Duran}, {Dainotti}  \& {Becerra}}{{Fraija}
  et~al.}{2020}]{2020ApJ...905..112F}
{Fraija} N.,  {Laskar} T.,  {Dichiara} S.,  {Beniamini} P.,  {Duran} R.~B.,
  {Dainotti} M.~G.,   {Becerra} R.~L.,  2020, \mn@doi [\apj]
  {10.3847/1538-4357/abc41a}, \href
  {https://ui.adsabs.harvard.edu/abs/2020ApJ...905..112F} {905, 112}

\bibitem[\protect\citeauthoryear{{Fraija}, {Dainotti}, {Ugale}, {Jyoti}  \&
  {Warren}}{{Fraija} et~al.}{2022}]{2022ApJ...934..188F}
{Fraija} N.,  {Dainotti} M.~G.,  {Ugale} S.,  {Jyoti} D.,   {Warren} D.~C.,
  2022, \mn@doi [\apj] {10.3847/1538-4357/ac7a9c}, \href
  {https://ui.adsabs.harvard.edu/abs/2022ApJ...934..188F} {934, 188}

\bibitem[\protect\citeauthoryear{{Fraija}, {Dainotti}, {Kamenetskaia}, {Levine}
   \& {Galvan-Gamez}}{{Fraija} et~al.}{2023a}]{2023MNRAS.525.1630F}
{Fraija} N.,  {Dainotti} M.~G.,  {Kamenetskaia} B.~B.,  {Levine} D.,
  {Galvan-Gamez} A.,  2023a, \mn@doi [\mnras] {10.1093/mnras/stad2236}, \href
  {https://ui.adsabs.harvard.edu/abs/2023MNRAS.525.1630F} {525, 1630}

\bibitem[\protect\citeauthoryear{{Fraija}, {Dainotti}, {Levine}, {Kamenetskaia}
   \& {Galvan-Gamez}}{{Fraija} et~al.}{2023b}]{2023ApJ...958..126F}
{Fraija} N.,  {Dainotti} M.~G.,  {Levine} D.,  {Kamenetskaia} B.~B.,
  {Galvan-Gamez} A.,  2023b, \mn@doi [\apj] {10.3847/1538-4357/acfb7f}, \href
  {https://ui.adsabs.harvard.edu/abs/2023ApJ...958..126F} {958, 126}

\bibitem[\protect\citeauthoryear{{Fraija} et~al.,}{{Fraija}
  et~al.}{2024a}]{2024MNRAS.527.1674F}
{Fraija} N.,  et~al., 2024a, \mn@doi [\mnras] {10.1093/mnras/stad3243}, \href
  {https://ui.adsabs.harvard.edu/abs/2024MNRAS.527.1674F} {527, 1674}

\bibitem[\protect\citeauthoryear{{Fraija}, {Dainotti}, {Betancourt
  Kamenetskaia}, {Galv{\'a}n-G{\'a}mez}  \& {Aguilar-Ruiz}}{{Fraija}
  et~al.}{2024b}]{2024MNRAS.527.1884F}
{Fraija} N.,  {Dainotti} M.~G.,  {Betancourt Kamenetskaia} B.,
  {Galv{\'a}n-G{\'a}mez} A.,   {Aguilar-Ruiz} E.,  2024b, \mn@doi [\mnras]
  {10.1093/mnras/stad3272}, \href
  {https://ui.adsabs.harvard.edu/abs/2024MNRAS.527.1884F} {527, 1884}

\bibitem[\protect\citeauthoryear{{Fryer}, {Woosley}  \& {Hartmann}}{{Fryer}
  et~al.}{1999}]{1999ApJ...526..152F}
{Fryer} C.~L.,  {Woosley} S.~E.,   {Hartmann} D.~H.,  1999, \mn@doi [\apj]
  {10.1086/307992}, \href
  {https://ui.adsabs.harvard.edu/abs/1999ApJ...526..152F} {526, 152}

\bibitem[\protect\citeauthoryear{Fryxell et~al.,}{Fryxell
  et~al.}{2000}]{fryxell2000flash}
Fryxell B.,  et~al., 2000, The Astrophysical Journal Supplement Series, 131,
  273

\bibitem[\protect\citeauthoryear{Gao, Ding, Wu, Dai  \& Zhang}{Gao
  et~al.}{2015}]{Gao2015GRB}
Gao H.,  Ding X.-H.,  Wu X.,  Dai Z.,   Zhang B.,  2015, \mn@doi [The
  Astrophysical Journal] {10.1088/0004-637X/807/2/163}, 807

\bibitem[\protect\citeauthoryear{Geppert, Page  \& Zannias}{Geppert
  et~al.}{1999}]{geppert1999submergence}
Geppert U.,  Page D.,   Zannias T.,  1999, Astronomy and Astrophysics, v. 345,
  p. 847-854 (1999), 345, 847

\bibitem[\protect\citeauthoryear{{Giacomazzo} \& {Perna}}{{Giacomazzo} \&
  {Perna}}{2013}]{2013ApJ...771L..26G}
{Giacomazzo} B.,  {Perna} R.,  2013, \mn@doi [\apjl]
  {10.1088/2041-8205/771/2/L26}, \href
  {https://ui.adsabs.harvard.edu/abs/2013ApJ...771L..26G} {771, L26}

\bibitem[\protect\citeauthoryear{{Goldstein} et~al.,}{{Goldstein}
  et~al.}{2017}]{2017ApJ...848L..14G}
{Goldstein} A.,  et~al., 2017, \mn@doi [\apjl] {10.3847/2041-8213/aa8f41},
  \href {http://adsabs.harvard.edu/abs/2017ApJ...848L..14G} {848, L14}

\bibitem[\protect\citeauthoryear{{Gottlieb}, {Nakar}, {Piran}  \&
  {Hotokezaka}}{{Gottlieb} et~al.}{2018}]{2018MNRAS.479..588G}
{Gottlieb} O.,  {Nakar} E.,  {Piran} T.,   {Hotokezaka} K.,  2018, \mn@doi
  [\mnras] {10.1093/mnras/sty1462}, \href
  {https://ui.adsabs.harvard.edu/abs/2018MNRAS.479..588G} {479, 588}

\bibitem[\protect\citeauthoryear{Gourgouliatos \& Cumming}{Gourgouliatos \&
  Cumming}{2014a}]{GourgouliatosCumming2014}
Gourgouliatos K.~N.,  Cumming A.,  2014a, Phys. Rev. Lett., 112, 171101

\bibitem[\protect\citeauthoryear{Gourgouliatos \& Cumming}{Gourgouliatos \&
  Cumming}{2014b}]{gourgouliatos2014hall}
Gourgouliatos K.,  Cumming A.,  2014b, Monthly Notices of the Royal
  Astronomical Society, 438, 1618

\bibitem[\protect\citeauthoryear{{Granot}, {K{\"o}nigl}  \& {Piran}}{{Granot}
  et~al.}{2006}]{2006MNRAS.370.1946G}
{Granot} J.,  {K{\"o}nigl} A.,   {Piran} T.,  2006, \mn@doi [\mnras]
  {10.1111/j.1365-2966.2006.10621.x}, \href
  {https://ui.adsabs.harvard.edu/abs/2006MNRAS.370.1946G} {370, 1946}

\bibitem[\protect\citeauthoryear{{Grossman}, {Korobkin}, {Rosswog}  \&
  {Piran}}{{Grossman} et~al.}{2014}]{2014MNRAS.439..757G}
{Grossman} D.,  {Korobkin} O.,  {Rosswog} S.,   {Piran} T.,  2014, \mn@doi
  [\mnras] {10.1093/mnras/stt2503}, \href
  {https://ui.adsabs.harvard.edu/abs/2014MNRAS.439..757G} {439, 757}

\bibitem[\protect\citeauthoryear{Hajela et~al.,}{Hajela
  et~al.}{2022a}]{hajela2022evidence}
Hajela A.,  et~al., 2022a, The Astrophysical Journal Letters, 927, L17

\bibitem[\protect\citeauthoryear{{Hajela} et~al.,}{{Hajela}
  et~al.}{2022b}]{2022ApJ...927L..17H}
{Hajela} A.,  et~al., 2022b, \mn@doi [\apjl] {10.3847/2041-8213/ac504a}, \href
  {https://ui.adsabs.harvard.edu/abs/2022ApJ...927L..17H} {927, L17}

\bibitem[\protect\citeauthoryear{Haskell, Samuelsson, Glampedakis  \&
  Andersson}{Haskell et~al.}{2008}]{haskell2008modelling}
Haskell B.,  Samuelsson L.,  Glampedakis K.,   Andersson N.,  2008, Monthly
  Notices of the Royal Astronomical Society, 385, 531

\bibitem[\protect\citeauthoryear{Ho}{Ho}{2011}]{art_ho}
Ho W.~C.,  2011, Monthly Notices of the Royal Astronomical Society, 414, 2567

\bibitem[\protect\citeauthoryear{{Hotokezaka}, {Kiuchi}, {Shibata}, {Nakar}  \&
  {Piran}}{{Hotokezaka} et~al.}{2018}]{2018ApJ...867...95H}
{Hotokezaka} K.,  {Kiuchi} K.,  {Shibata} M.,  {Nakar} E.,   {Piran} T.,  2018,
  \mn@doi [\apj] {10.3847/1538-4357/aadf92}, \href
  {http://adsabs.harvard.edu/abs/2018ApJ...867...95H} {867, 95}

\bibitem[\protect\citeauthoryear{{Hurley} et~al.,}{{Hurley}
  et~al.}{2005}]{2005Natur.434.1098H}
{Hurley} K.,  et~al., 2005, \mn@doi [\nat] {10.1038/nature03519}, \href
  {https://ui.adsabs.harvard.edu/abs/2005Natur.434.1098H} {434, 1098}

\bibitem[\protect\citeauthoryear{{Ioka}}{{Ioka}}{2005}]{2005PThPh.114.1317I}
{Ioka} K.,  2005, \mn@doi [Progress of Theoretical Physics]
  {10.1143/PTP.114.1317}, \href
  {https://ui.adsabs.harvard.edu/abs/2005PThPh.114.1317I} {114, 1317}

\bibitem[\protect\citeauthoryear{{Ioka}, {Toma}, {Yamazaki}  \&
  {Nakamura}}{{Ioka} et~al.}{2006}]{2006A&A...458....7I}
{Ioka} K.,  {Toma} K.,  {Yamazaki} R.,   {Nakamura} T.,  2006, \mn@doi [\aap]
  {10.1051/0004-6361:20064939}, \href
  {https://ui.adsabs.harvard.edu/abs/2006A&A...458....7I} {458, 7}

\bibitem[\protect\citeauthoryear{{Kasliwal} et~al.,}{{Kasliwal}
  et~al.}{2017}]{2017Sci...358.1559K}
{Kasliwal} M.~M.,  et~al., 2017, \mn@doi [Science] {10.1126/science.aap9455},
  \href {http://adsabs.harvard.edu/abs/2017Sci...358.1559K} {358, 1559}

\bibitem[\protect\citeauthoryear{{Kisaka}, {Ioka}, {Kashiyama}  \&
  {Nakamura}}{{Kisaka} et~al.}{2017}]{2017arXiv171100243K}
{Kisaka} S.,  {Ioka} K.,  {Kashiyama} K.,   {Nakamura} T.,  2017, preprint,
  \href {http://adsabs.harvard.edu/abs/2017arXiv171100243K} {} (\mn@eprint
  {arXiv} {1711.00243})

\bibitem[\protect\citeauthoryear{{Kobayashi}}{{Kobayashi}}{2000}]{2000ApJ...545..807K}
{Kobayashi} S.,  2000, \mn@doi [\apj] {10.1086/317869}, \href
  {http://adsabs.harvard.edu/abs/2000ApJ...545..807K} {545, 807}

\bibitem[\protect\citeauthoryear{Kojima \& Kisaka}{Kojima \&
  Kisaka}{2012}]{kojima2012magnetic}
Kojima Y.,  Kisaka S.,  2012, Monthly Notices of the Royal Astronomical
  Society, 421, 2722

\bibitem[\protect\citeauthoryear{Konar}{Konar}{2017}]{konar2017magnetic}
Konar S.,  2017, Journal of Astrophysics and Astronomy, 38, 47

\bibitem[\protect\citeauthoryear{Konar \& Choudhuri}{Konar \&
  Choudhuri}{2002}]{konar2002diamagnetic}
Konar S.,  Choudhuri A.,  2002, in 34th COSPAR Scientific Assembly. p.~721

\bibitem[\protect\citeauthoryear{Kong, Wong, Huang  \& Cheng}{Kong
  et~al.}{2010}]{10.1111/j.1365-2966.2009.15886.x}
Kong S.~W.,  Wong A. Y.~L.,  Huang Y.~F.,   Cheng K.~S.,  2010, \mn@doi
  [Monthly Notices of the Royal Astronomical Society]
  {10.1111/j.1365-2966.2009.15886.x}, 402, 409

\bibitem[\protect\citeauthoryear{{Kumar} \& {Panaitescu}}{{Kumar} \&
  {Panaitescu}}{2003}]{2003MNRAS.346..905K}
{Kumar} P.,  {Panaitescu} A.,  2003, \mn@doi [\mnras]
  {10.1111/j.1365-2966.2003.07138.x}, \href
  {http://adsabs.harvard.edu/abs/2003MNRAS.346..905K} {346, 905}

\bibitem[\protect\citeauthoryear{{Kumar} \& {Zhang}}{{Kumar} \&
  {Zhang}}{2015}]{2015PhR...561....1K}
{Kumar} P.,  {Zhang} B.,  2015, \mn@doi [\physrep]
  {10.1016/j.physrep.2014.09.008}, \href
  {http://adsabs.harvard.edu/abs/2015PhR...561....1K} {561, 1}

\bibitem[\protect\citeauthoryear{{Kumar}, {Narayan}  \& {Johnson}}{{Kumar}
  et~al.}{2008}]{Kumar2008}
{Kumar} P.,  {Narayan} R.,   {Johnson} J.~L.,  2008, \mn@doi [Science]
  {10.1126/science.1159003}, \href
  {http://adsabs.harvard.edu/abs/2008Sci...321..376K} {321, 376}

\bibitem[\protect\citeauthoryear{{Lamb} \& {Kobayashi}}{{Lamb} \&
  {Kobayashi}}{2017}]{2017MNRAS.472.4953L}
{Lamb} G.~P.,  {Kobayashi} S.,  2017, \mn@doi [\mnras] {10.1093/mnras/stx2345},
  \href {http://adsabs.harvard.edu/abs/2017MNRAS.472.4953L} {472, 4953}

\bibitem[\protect\citeauthoryear{{Lazzati}, {L{\'o}pez-C{\'a}mara},
  {Cantiello}, {Morsony}, {Perna}  \& {Workman}}{{Lazzati}
  et~al.}{2017}]{2017ApJ...848L...6L}
{Lazzati} D.,  {L{\'o}pez-C{\'a}mara} D.,  {Cantiello} M.,  {Morsony} B.~J.,
  {Perna} R.,   {Workman} J.~C.,  2017, \mn@doi [\apjl]
  {10.3847/2041-8213/aa8f3d}, \href
  {http://adsabs.harvard.edu/abs/2017ApJ...848L...6L} {848, L6}

\bibitem[\protect\citeauthoryear{{Lemoine}}{{Lemoine}}{2013}]{2013MNRAS.428..845L}
{Lemoine} M.,  2013, \mn@doi [\mnras] {10.1093/mnras/sts081}, \href
  {https://ui.adsabs.harvard.edu/abs/2013MNRAS.428..845L} {428, 845}

\bibitem[\protect\citeauthoryear{Levin \& Lyutikov}{Levin \&
  Lyutikov}{2012}]{levin2012dynamics}
Levin Y.,  Lyutikov M.,  2012, Monthly Notices of the Royal Astronomical
  Society, 427, 1574

\bibitem[\protect\citeauthoryear{Li \& Beloborodov}{Li \&
  Beloborodov}{2015a}]{LiBeloborodov2015}
Li S.,  Beloborodov A.~M.,  2015a, Astrophys. J., 815, 25

\bibitem[\protect\citeauthoryear{Li \& Beloborodov}{Li \&
  Beloborodov}{2015b}]{li2015plastic}
Li X.,  Beloborodov A.~M.,  2015b, The Astrophysical Journal, 815, 25

\bibitem[\protect\citeauthoryear{{Li} \& {Dai}}{{Li} \&
  {Dai}}{2021}]{2021ApJ...918...52L}
{Li} L.,  {Dai} Z.-G.,  2021, \mn@doi [\apj] {10.3847/1538-4357/ac0974}, \href
  {https://ui.adsabs.harvard.edu/abs/2021ApJ...918...52L} {918, 52}

\bibitem[\protect\citeauthoryear{Li, Wu, Lei, Dai, Liang  \& Ryde}{Li
  et~al.}{2018}]{li18}
Li L.,  Wu X.-F.,  Lei W.-H.,  Dai Z.-G.,  Liang E.-W.,   Ryde F.,  2018, The
  Astrophysical Journal Supplement Series, 236, 26

\bibitem[\protect\citeauthoryear{{Lien} et~al.,}{{Lien}
  et~al.}{2016}]{2016ApJ...829....7L}
{Lien} A.,  et~al., 2016, \mn@doi [\apj] {10.3847/0004-637X/829/1/7}, \href
  {https://ui.adsabs.harvard.edu/abs/2016ApJ...829....7L} {829, 7}

\bibitem[\protect\citeauthoryear{{L{\"u}} \& {Zhang}}{{L{\"u}} \&
  {Zhang}}{2014a}]{2014ApJ...785...74L}
{L{\"u}} H.-J.,  {Zhang} B.,  2014a, \mn@doi [\apj]
  {10.1088/0004-637X/785/1/74}, \href
  {https://ui.adsabs.harvard.edu/abs/2014ApJ...785...74L} {785, 74}

\bibitem[\protect\citeauthoryear{L{\"u} \& Zhang}{L{\"u} \&
  Zhang}{2014b}]{lu14}
L{\"u} H.-J.,  Zhang B.,  2014b, The Astrophysical Journal, 785, 74

\bibitem[\protect\citeauthoryear{{MacFadyen} \& {Woosley}}{{MacFadyen} \&
  {Woosley}}{1999a}]{Macfadyen+99col}
{MacFadyen} A.~I.,  {Woosley} S.~E.,  1999a, \mn@doi [\apj] {10.1086/307790},
  \href {http://adsabs.harvard.edu/abs/1999ApJ...524..262M} {524, 262}

\bibitem[\protect\citeauthoryear{{MacFadyen} \& {Woosley}}{{MacFadyen} \&
  {Woosley}}{1999b}]{1999ApJ...524..262M}
{MacFadyen} A.~I.,  {Woosley} S.~E.,  1999b, \mn@doi [\apj] {10.1086/307790},
  \href {https://ui.adsabs.harvard.edu/abs/1999ApJ...524..262M} {524, 262}

\bibitem[\protect\citeauthoryear{{Margalit} \& {Piran}}{{Margalit} \&
  {Piran}}{2020}]{2020arXiv200413028M}
{Margalit} B.,  {Piran} T.,  2020, arXiv e-prints, \href
  {https://ui.adsabs.harvard.edu/abs/2020arXiv200413028M} {p. arXiv:2004.13028}

\bibitem[\protect\citeauthoryear{{Margutti} et~al.,}{{Margutti}
  et~al.}{2017}]{2017ApJ...848L..20M}
{Margutti} R.,  et~al., 2017, \mn@doi [\apjl] {10.3847/2041-8213/aa9057}, \href
  {http://cdsads.u-strasbg.fr/abs/2017ApJ...848L..20M} {848, L20}

\bibitem[\protect\citeauthoryear{Metzger, Giannios, Thompson, Bucciantini  \&
  Quataert}{Metzger et~al.}{2011b}]{metzger2011protomagnetar}
Metzger B.,  Giannios D.,  Thompson T.,  Bucciantini N.,   Quataert E.,  2011b,
  Monthly Notices of the Royal Astronomical Society, 413, 2031

\bibitem[\protect\citeauthoryear{{Metzger}, {Giannios}, {Thompson},
  {Bucciantini}  \& {Quataert}}{{Metzger} et~al.}{2011a}]{2011MNRAS.413.2031M}
{Metzger} B.~D.,  {Giannios} D.,  {Thompson} T.~A.,  {Bucciantini} N.,
  {Quataert} E.,  2011a, \mn@doi [\mnras] {10.1111/j.1365-2966.2011.18280.x},
  \href {https://ui.adsabs.harvard.edu/abs/2011MNRAS.413.2031M} {413, 2031}

\bibitem[\protect\citeauthoryear{Metzger, Beniamini  \& Giannios}{Metzger
  et~al.}{2018a}]{Metzger2018Effects}
Metzger B.,  Beniamini P.,   Giannios D.,  2018a, \mn@doi [The Astrophysical
  Journal] {10.3847/1538-4357/aab70c}, 857

\bibitem[\protect\citeauthoryear{{Metzger}, {Beniamini}  \&
  {Giannios}}{{Metzger} et~al.}{2018b}]{2018ApJ...857...95M}
{Metzger} B.~D.,  {Beniamini} P.,   {Giannios} D.,  2018b, \mn@doi [\apj]
  {10.3847/1538-4357/aab70c}, \href
  {https://ui.adsabs.harvard.edu/abs/2018ApJ...857...95M} {857, 95}

\bibitem[\protect\citeauthoryear{Michel}{Michel}{1994}]{michel1994magnetic}
Michel F.~C.,  1994, The Astrophysical Journal, Part 1, vol. 431, no. 1, p.
  397-401, 431, 397

\bibitem[\protect\citeauthoryear{{Moderski}, {Sikora}  \& {Bulik}}{{Moderski}
  et~al.}{2000}]{2000ApJ...529..151M}
{Moderski} R.,  {Sikora} M.,   {Bulik} T.,  2000, \mn@doi [\apj]
  {10.1086/308257}, \href
  {https://ui.adsabs.harvard.edu/abs/2000ApJ...529..151M} {529, 151}

\bibitem[\protect\citeauthoryear{{Mukherjee}}{{Mukherjee}}{2017a}]{2017JApA...38...48M}
{Mukherjee} D.,  2017a, \mn@doi [Journal of Astrophysics and Astronomy]
  {10.1007/s12036-017-9465-6}, \href
  {https://ui.adsabs.harvard.edu/abs/2017JApA...38...48M} {38, 48}

\bibitem[\protect\citeauthoryear{Mukherjee}{Mukherjee}{2017b}]{mukherjee2017revisiting}
Mukherjee D.,  2017b, Journal of Astrophysics and Astronomy, 38, 48

\bibitem[\protect\citeauthoryear{Muslimov \& Page}{Muslimov \&
  Page}{1995}]{art_dany}
Muslimov A.,  Page D.,  1995, The Astrophysical Journal, 458, 347

\bibitem[\protect\citeauthoryear{{Narayan}, {Paczynski}  \& {Piran}}{{Narayan}
  et~al.}{1992}]{1992ApJ...395L..83N}
{Narayan} R.,  {Paczynski} B.,   {Piran} T.,  1992, \mn@doi [\apjl]
  {10.1086/186493}, \href {http://adsabs.harvard.edu/abs/1992ApJ...395L..83N}
  {395, L83}

\bibitem[\protect\citeauthoryear{{Nedora}, {Radice}, {Bernuzzi}, {Perego},
  {Daszuta}, {Endrizzi}, {Prakash}  \& {Schianchi}}{{Nedora}
  et~al.}{2021}]{2021MNRAS.506.5908N}
{Nedora} V.,  {Radice} D.,  {Bernuzzi} S.,  {Perego} A.,  {Daszuta} B.,
  {Endrizzi} A.,  {Prakash} A.,   {Schianchi} F.,  2021, \mn@doi [\mnras]
  {10.1093/mnras/stab2004}, \href
  {https://ui.adsabs.harvard.edu/abs/2021MNRAS.506.5908N} {506, 5908}

\bibitem[\protect\citeauthoryear{Negreiros \& Bernal}{Negreiros \&
  Bernal}{2015}]{negreiros2015growth}
Negreiros R.,  Bernal C.,  2015, Growth of the Magnetic Field in Young Neutron
  Stars (\mn@eprint {arXiv} {1505.02823})

\bibitem[\protect\citeauthoryear{Newville, Stensitzki, Allen  et~al.}{Newville
  et~al.}{2014}]{newville_matthew_2014_11813}
Newville M.,  Stensitzki T.,  Allen D.~B.,   et~al., 2014, {LMFIT: Non-Linear
  Least-Square Minimization and Curve-Fitting for Python},
  \mn@doi{10.5281/zenodo.11813}, \url {https://doi.org/10.5281/zenodo.11813}

\bibitem[\protect\citeauthoryear{{Nousek} et~al.,}{{Nousek}
  et~al.}{2006}]{2006ApJ...642..389N}
{Nousek} J.~A.,  et~al., 2006, \mn@doi [\apj] {10.1086/500724}, \href
  {https://ui.adsabs.harvard.edu/abs/2006ApJ...642..389N} {642, 389}

\bibitem[\protect\citeauthoryear{{Oates}, {Page}, {De Pasquale}, {Schady},
  {Breeveld}, {Holland}, {Kuin}  \& {Marshall}}{{Oates}
  et~al.}{2012}]{oates2012}
{Oates} S.~R.,  {Page} M.~J.,  {De Pasquale} M.,  {Schady} P.,  {Breeveld}
  A.~A.,  {Holland} S.~T.,  {Kuin} N.~P.~M.,   {Marshall} F.~E.,  2012, \mn@doi
  [\mnras] {10.1111/j.1745-3933.2012.01331.x}, \href
  {http://adsabs.harvard.edu/abs/2012MNRAS.426L..86O} {426, L86}

\bibitem[\protect\citeauthoryear{{{\"O}zel} \& {Freire}}{{{\"O}zel} \&
  {Freire}}{2016}]{2016ARA&A..54..401O}
{{\"O}zel} F.,  {Freire} P.,  2016, \mn@doi [\araa]
  {10.1146/annurev-astro-081915-023322}, \href
  {https://ui.adsabs.harvard.edu/abs/2016ARA&A..54..401O} {54, 401}

\bibitem[\protect\citeauthoryear{{Paczy{\'n}ski}}{{Paczy{\'n}ski}}{1998}]{1998ApJ...494L..45P}
{Paczy{\'n}ski} B.,  1998, \mn@doi [\apjl] {10.1086/311148}, \href
  {https://ui.adsabs.harvard.edu/abs/1998ApJ...494L..45P} {494, L45}

\bibitem[\protect\citeauthoryear{{Panaitescu} \& {Kumar}}{{Panaitescu} \&
  {Kumar}}{2002}]{2002ApJ...571..779P}
{Panaitescu} A.,  {Kumar} P.,  2002, \mn@doi [\apj] {10.1086/340094}, \href
  {http://adsabs.harvard.edu/abs/2002ApJ...571..779P} {571, 779}

\bibitem[\protect\citeauthoryear{{Panaitescu} \&
  {M{\'e}sz{\'a}ros}}{{Panaitescu} \&
  {M{\'e}sz{\'a}ros}}{1998}]{1998ApJ...501..772P}
{Panaitescu} A.,  {M{\'e}sz{\'a}ros} P.,  1998, \mn@doi [\apj]
  {10.1086/305856}, \href {http://adsabs.harvard.edu/abs/1998ApJ...501..772P}
  {501, 772}

\bibitem[\protect\citeauthoryear{{Panaitescu}, {M{\'e}sz{\'a}ros}, {Burrows},
  {Nousek}, {Gehrels}, {O'Brien}  \& {Willingale}}{{Panaitescu}
  et~al.}{2006}]{2006MNRAS.369.2059P}
{Panaitescu} A.,  {M{\'e}sz{\'a}ros} P.,  {Burrows} D.,  {Nousek} J.,
  {Gehrels} N.,  {O'Brien} P.,   {Willingale} R.,  2006, \mn@doi [\mnras]
  {10.1111/j.1365-2966.2006.10453.x}, \href
  {https://ui.adsabs.harvard.edu/abs/2006MNRAS.369.2059P} {369, 2059}

\bibitem[\protect\citeauthoryear{Payne \& Melatos}{Payne \&
  Melatos}{2004}]{10.1111/j.1365-2966.2004.07798.x}
Payne D. J.~B.,  Melatos A.,  2004, \mn@doi [Monthly Notices of the Royal
  Astronomical Society] {10.1111/j.1365-2966.2004.07798.x}, 351, 569

\bibitem[\protect\citeauthoryear{Perna \& Pons}{Perna \&
  Pons}{2011}]{Perna2011AUM}
Perna R.,  Pons J.~A.,  2011, The Astrophysical Journal Letters, 727

\bibitem[\protect\citeauthoryear{{Piro} \& {Ott}}{{Piro} \&
  {Ott}}{2011a}]{2011ApJ...736..108P}
{Piro} A.~L.,  {Ott} C.~D.,  2011a, \mn@doi [\apj]
  {10.1088/0004-637X/736/2/108}, \href
  {https://ui.adsabs.harvard.edu/abs/2011ApJ...736..108P} {736, 108}

\bibitem[\protect\citeauthoryear{Piro \& Ott}{Piro \&
  Ott}{2011b}]{piro2011supernova}
Piro A.~L.,  Ott C.~D.,  2011b, The Astrophysical Journal, 736, 108

\bibitem[\protect\citeauthoryear{Pons \& Geppert}{Pons \&
  Geppert}{2007}]{pons2007magnetic}
Pons J.~A.,  Geppert U.,  2007, Astronomy \& Astrophysics, 470, 303

\bibitem[\protect\citeauthoryear{Pons \& Perna}{Pons \&
  Perna}{2011}]{pons2011magnetars}
Pons J.~A.,  Perna R.,  2011, The Astrophysical Journal, 741, 123

\bibitem[\protect\citeauthoryear{{Price} \& {Rosswog}}{{Price} \&
  {Rosswog}}{2006}]{2006Sci...312..719P}
{Price} D.~J.,  {Rosswog} S.,  2006, \mn@doi [Science]
  {10.1126/science.1125201}, \href
  {https://ui.adsabs.harvard.edu/abs/2006Sci...312..719P} {312, 719}

\bibitem[\protect\citeauthoryear{Quataert \& Kasen}{Quataert \&
  Kasen}{2012b}]{quataert2012swift}
Quataert E.,  Kasen D.,  2012b, Monthly Notices of the Royal Astronomical
  Society: Letters, 419, L1

\bibitem[\protect\citeauthoryear{{Quataert} \& {Kasen}}{{Quataert} \&
  {Kasen}}{2012a}]{2012MNRAS.419L...1Q}
{Quataert} E.,  {Kasen} D.,  2012a, \mn@doi [\mnras]
  {10.1111/j.1745-3933.2011.01151.x}, \href
  {https://ui.adsabs.harvard.edu/abs/2012MNRAS.419L...1Q} {419, L1}

\bibitem[\protect\citeauthoryear{{Racusin} et~al.,}{{Racusin}
  et~al.}{2009}]{2009ApJ...698...43R}
{Racusin} J.~L.,  et~al., 2009, \mn@doi [\apj] {10.1088/0004-637X/698/1/43},
  \href {http://adsabs.harvard.edu/abs/2009ApJ...698...43R} {698, 43}

\bibitem[\protect\citeauthoryear{Ren \& Dai}{Ren \&
  Dai}{2022}]{10.1093/mnras/stac797}
Ren J.,  Dai Z.~G.,  2022, \mn@doi [Monthly Notices of the Royal Astronomical
  Society] {10.1093/mnras/stac797}, 512, 5572

\bibitem[\protect\citeauthoryear{Riley}{Riley}{2021}]{Riley2021}
Riley T. E. e.~a.,  2021, Astrophys. J. Lett., 918, L27

\bibitem[\protect\citeauthoryear{Roberts et~al.,}{Roberts
  et~al.}{2021}]{Roberts2021Rapid}
Roberts O.,  et~al., 2021, \mn@doi [Nature] {10.1038/s41586-020-03077-8}, 589,
  207

\bibitem[\protect\citeauthoryear{Rogers \& Safi-Harb}{Rogers \&
  Safi-Harb}{2016}]{Rogers_2016}
Rogers A.,  Safi-Harb S.,  2016, \mn@doi [Monthly Notices of the Royal
  Astronomical Society] {10.1093/mnras/stw014}, 457, 1180–1189

\bibitem[\protect\citeauthoryear{Rowlinson}{Rowlinson}{2013}]{Rowlinson2013}
Rowlinson A. e.~a.,  2013, Mon. Not. R. Astron. Soc., 430, 1061

\bibitem[\protect\citeauthoryear{{Rowlinson}, {O'Brien}, {Tanvir}, {Zhang},
  {Evans}  \& {et al.}}{{Rowlinson} et~al.}{2010}]{2010MNRAS.409..531R}
{Rowlinson} A.,  {O'Brien} P.~T.,  {Tanvir} N.~R.,  {Zhang} B.,  {Evans} P.~A.,
    {et al.} 2010, \mn@doi [\mnras] {10.1111/j.1365-2966.2010.17354.x}, \href
  {https://ui.adsabs.harvard.edu/abs/2010MNRAS.409..531R} {409, 531}

\bibitem[\protect\citeauthoryear{{Rowlinson}, {O'Brien}, {Metzger}, {Tanvir}
  \& {Levan}}{{Rowlinson} et~al.}{2013a}]{2013MNRAS.430.1061R}
{Rowlinson} A.,  {O'Brien} P.~T.,  {Metzger} B.~D.,  {Tanvir} N.~R.,   {Levan}
  A.~J.,  2013a, \mn@doi [\mnras] {10.1093/mnras/sts683}, \href
  {https://ui.adsabs.harvard.edu/abs/2013MNRAS.430.1061R} {430, 1061}

\bibitem[\protect\citeauthoryear{Rowlinson, O'brien, Metzger, Tanvir  \&
  Levan}{Rowlinson et~al.}{2013b}]{rowlinson2013signatures}
Rowlinson A.,  O'brien P.,  Metzger B.,  Tanvir N.,   Levan A.~J.,  2013b,
  Monthly Notices of the Royal Astronomical Society, 430, 1061

\bibitem[\protect\citeauthoryear{{Ryan}, {van Eerten}, {Troja}, {Piro},
  {O'Connor}  \& {Ricci}}{{Ryan} et~al.}{2024}]{2024ApJ...975..131R}
{Ryan} G.,  {van Eerten} H.,  {Troja} E.,  {Piro} L.,  {O'Connor} B.,   {Ricci}
  R.,  2024, \mn@doi [\apj] {10.3847/1538-4357/ad6a14}, \href
  {https://ui.adsabs.harvard.edu/abs/2024ApJ...975..131R} {975, 131}

\bibitem[\protect\citeauthoryear{{Sari} \& {M{\'e}sz{\'a}ros}}{{Sari} \&
  {M{\'e}sz{\'a}ros}}{2000}]{2000ApJ...535L..33S}
{Sari} R.,  {M{\'e}sz{\'a}ros} P.,  2000, \mn@doi [\apjl] {10.1086/312689},
  \href {http://adsabs.harvard.edu/abs/2000ApJ...535L..33S} {535, L33}

\bibitem[\protect\citeauthoryear{{Sari} \& {Piran}}{{Sari} \&
  {Piran}}{1995}]{1995ApJ...455L.143S}
{Sari} R.,  {Piran} T.,  1995, \mn@doi [\apjl] {10.1086/309835}, \href
  {http://adsabs.harvard.edu/abs/1995ApJ...455L.143S} {455, L143}

\bibitem[\protect\citeauthoryear{Sarin, Lasky  \& Ashton}{Sarin
  et~al.}{2020}]{Sarin2020Interpreting}
Sarin N.,  Lasky P.,   Ashton G.,  2020, \mn@doi [Monthly Notices of the Royal
  Astronomical Society] {10.1093/mnras/staa3090}, 499, 5986

\bibitem[\protect\citeauthoryear{{Savchenko} et~al.,}{{Savchenko}
  et~al.}{2017}]{2017ApJ...848L..15S}
{Savchenko} V.,  et~al., 2017, \mn@doi [\apjl] {10.3847/2041-8213/aa8f94},
  \href {http://adsabs.harvard.edu/abs/2017ApJ...848L..15S} {848, L15}

\bibitem[\protect\citeauthoryear{Shabaltas \& Lai}{Shabaltas \&
  Lai}{2012}]{Shabaltas_2012}
Shabaltas N.,  Lai D.,  2012, \mn@doi [The Astrophysical Journal]
  {10.1088/0004-637x/748/2/148}, 748, 148

\bibitem[\protect\citeauthoryear{Shalybkov \& Urpin}{Shalybkov \&
  Urpin}{1995}]{shalybkov1995ambipolar}
Shalybkov D.,  Urpin V.,  1995, Monthly Notices of the Royal Astronomical
  Society, 273, 643

\bibitem[\protect\citeauthoryear{{Sironi} \& {Giannios}}{{Sironi} \&
  {Giannios}}{2013}]{2013ApJ...778..107S}
{Sironi} L.,  {Giannios} D.,  2013, \mn@doi [\apj]
  {10.1088/0004-637X/778/2/107}, \href
  {https://ui.adsabs.harvard.edu/abs/2013ApJ...778..107S} {778, 107}

\bibitem[\protect\citeauthoryear{Skiathas \& Gourgouliatos}{Skiathas \&
  Gourgouliatos}{2024}]{skiathas2024combined}
Skiathas D.,  Gourgouliatos K.~N.,  2024, Monthly Notices of the Royal
  Astronomical Society, 528, 5178

\bibitem[\protect\citeauthoryear{{Srinivasaragavan}, {Dainotti}, {Fraija},
  {Hernandez}, {Nagataki}, {Lenart}, {Bowden}  \& {Wagner}}{{Srinivasaragavan}
  et~al.}{2020}]{Srinivasaragavan2020ApJ}
{Srinivasaragavan} G.~P.,  {Dainotti} M.~G.,  {Fraija} N.,  {Hernandez} X.,
  {Nagataki} S.,  {Lenart} A.,  {Bowden} L.,   {Wagner} R.,  2020, \mn@doi
  [\apj] {10.3847/1538-4357/abb702}, \href
  {https://ui.adsabs.harvard.edu/abs/2020ApJ...903...18S} {903, 18}

\bibitem[\protect\citeauthoryear{{Stella}, {Dall'Osso}, {Israel}  \&
  {Vecchio}}{{Stella} et~al.}{2005}]{2005ApJ...634L.165S}
{Stella} L.,  {Dall'Osso} S.,  {Israel} G.~L.,   {Vecchio} A.,  2005, \mn@doi
  [\apjl] {10.1086/498685}, \href
  {https://ui.adsabs.harvard.edu/abs/2005ApJ...634L.165S} {634, L165}

\bibitem[\protect\citeauthoryear{Suvorov \& Kokkotas}{Suvorov \&
  Kokkotas}{2020}]{SuvorovKokkotas2020}
Suvorov A.~G.,  Kokkotas K.~D.,  2020, Astrophys. J. Lett., 899, L10

\bibitem[\protect\citeauthoryear{Suvorov \& Melatos}{Suvorov \&
  Melatos}{2020}]{suvorov2020recycled}
Suvorov A.~G.,  Melatos A.,  2020, Monthly Notices of the Royal Astronomical
  Society, 499, 3243

\bibitem[\protect\citeauthoryear{{Tak}, {Omodei}, {Uhm}, {Racusin}, {Asano}  \&
  {McEnery}}{{Tak} et~al.}{2019}]{2019ApJ...883..134T}
{Tak} D.,  {Omodei} N.,  {Uhm} Z.~L.,  {Racusin} J.,  {Asano} K.,   {McEnery}
  J.,  2019, \mn@doi [\apj] {10.3847/1538-4357/ab3982}, \href
  {https://ui.adsabs.harvard.edu/abs/2019ApJ...883..134T} {883, 134}

\bibitem[\protect\citeauthoryear{{Tan}, {Matzner}  \& {McKee}}{{Tan}
  et~al.}{2001}]{2001ApJ...551..946T}
{Tan} J.~C.,  {Matzner} C.~D.,   {McKee} C.~F.,  2001, \mn@doi [\apj]
  {10.1086/320245}, \href {http://adsabs.harvard.edu/abs/2001ApJ...551..946T}
  {551, 946}

\bibitem[\protect\citeauthoryear{{Taylor}, {Frail}, {Berger}  \&
  {Kulkarni}}{{Taylor} et~al.}{2004}]{2004ApJ...609L...1T}
{Taylor} G.~B.,  {Frail} D.~A.,  {Berger} E.,   {Kulkarni} S.~R.,  2004,
  \mn@doi [\apjl] {10.1086/422554}, \href
  {http://adsabs.harvard.edu/abs/2004ApJ...609L...1T} {609, L1}

\bibitem[\protect\citeauthoryear{{Thompson}}{{Thompson}}{1994}]{1994MNRAS.270..480T}
{Thompson} C.,  1994, \mn@doi [\mnras] {10.1093/mnras/270.3.480}, \href
  {http://ukads.nottingham.ac.uk/abs/1994MNRAS.270..480T} {270, 480}

\bibitem[\protect\citeauthoryear{{Thompson} \& {Duncan}}{{Thompson} \&
  {Duncan}}{1993}]{1993ApJ...408..194T}
{Thompson} C.,  {Duncan} R.~C.,  1993, \mn@doi [\apj] {10.1086/172580}, \href
  {https://ui.adsabs.harvard.edu/abs/1993ApJ...408..194T} {408, 194}

\bibitem[\protect\citeauthoryear{{Torres-Forn{\'e}}, {Cerd{\'a}-Dur{\'a}n},
  {Pons}  \& {Font}}{{Torres-Forn{\'e}} et~al.}{2016}]{2016MNRAS.456.3813T}
{Torres-Forn{\'e}} A.,  {Cerd{\'a}-Dur{\'a}n} P.,  {Pons} J.~A.,   {Font}
  J.~A.,  2016, \mn@doi [\mnras] {10.1093/mnras/stv2926}, \href
  {https://ui.adsabs.harvard.edu/abs/2016MNRAS.456.3813T} {456, 3813}

\bibitem[\protect\citeauthoryear{Torres-Forné}{Torres-Forné}{2016}]{TorresForne2016}
Torres-Forné A. e.~a.,  2016, Mon. Not. R. Astron. Soc., 456, 3813

\bibitem[\protect\citeauthoryear{{Troja}, {Cusumano}, {O'Brien}, {Zhang},
  {Sbarufatti}  \& {et al.}}{{Troja} et~al.}{2007}]{2007ApJ...665..599T}
{Troja} E.,  {Cusumano} G.,  {O'Brien} P.~T.,  {Zhang} B.,  {Sbarufatti} B.,
  {et al.} 2007, \mn@doi [\apj] {10.1086/519450}, \href
  {https://ui.adsabs.harvard.edu/abs/2007ApJ...665..599T} {665, 599}

\bibitem[\protect\citeauthoryear{Troja, Piro, van Eerten  \& et al.}{Troja
  et~al.}{2017}]{troja2017a}
Troja E.,  Piro L.,  van Eerten H.,   et al. 2017, \mn@doi [Nature]
  {10.1038/nature24290}, 000, 1

\bibitem[\protect\citeauthoryear{{Urrutia}, {De Colle}, {Murguia-Berthier}  \&
  {Ramirez-Ruiz}}{{Urrutia} et~al.}{2021}]{2021MNRAS.503.4363U}
{Urrutia} G.,  {De Colle} F.,  {Murguia-Berthier} A.,   {Ramirez-Ruiz} E.,
  2021, \mn@doi [\mnras] {10.1093/mnras/stab723}, \href
  {https://ui.adsabs.harvard.edu/abs/2021MNRAS.503.4363U} {503, 4363}

\bibitem[\protect\citeauthoryear{{Usov}}{{Usov}}{1992}]{1992Natur.357..472U}
{Usov} V.~V.,  1992, \mn@doi [\nat] {10.1038/357472a0}, \href
  {http://adsabs.harvard.edu/abs/1992Natur.357..472U} {357, 472}

\bibitem[\protect\citeauthoryear{Vigan{\`o} \& Pons}{Vigan{\`o} \&
  Pons}{2012a}]{art_vigano}
Vigan{\`o} D.,  Pons J.~A.,  2012a, Monthly Notices of the Royal Astronomical
  Society, 425, 2487

\bibitem[\protect\citeauthoryear{{Vigan{\`o}} \& {Pons}}{{Vigan{\`o}} \&
  {Pons}}{2012b}]{2012MNRAS.425.2487V}
{Vigan{\`o}} D.,  {Pons} J.~A.,  2012b, \mn@doi [\mnras]
  {10.1111/j.1365-2966.2012.21679.x}, \href
  {https://ui.adsabs.harvard.edu/abs/2012MNRAS.425.2487V} {425, 2487}

\bibitem[\protect\citeauthoryear{Viganò, Rea, Pons, Perna, Aguilera  \&
  Miralles}{Viganò et~al.}{2013}]{Vigano2013}
Viganò D.,  Rea N.,  Pons J.~A.,  Perna R.,  Aguilera D.~N.,   Miralles J.~A.,
   2013, Monthly Notices of the Royal Astronomical Society, 434, 123

\bibitem[\protect\citeauthoryear{Vigelius \& Melatos}{Vigelius \&
  Melatos}{2009}]{vigelius2009resistive}
Vigelius M.,  Melatos A.,  2009, Monthly Notices of the Royal Astronomical
  Society, 395, 1985

\bibitem[\protect\citeauthoryear{{Woosley}}{{Woosley}}{1993}]{1993ApJ...405..273W}
{Woosley} S.~E.,  1993, \mn@doi [\apj] {10.1086/172359}, \href
  {https://ui.adsabs.harvard.edu/abs/1993ApJ...405..273W} {405, 273}

\bibitem[\protect\citeauthoryear{{Woosley} \& {Bloom}}{{Woosley} \&
  {Bloom}}{2006}]{Woosley2006ARA&A}
{Woosley} S.~E.,  {Bloom} J.~S.,  2006, \mn@doi [\araa]
  {10.1146/annurev.astro.43.072103.150558}, \href
  {https://ui.adsabs.harvard.edu/abs/2006ARA&A..44..507W} {44, 507}

\bibitem[\protect\citeauthoryear{{Woosley} \& {Heger}}{{Woosley} \&
  {Heger}}{2012}]{2012ApJ...752...32W}
{Woosley} S.~E.,  {Heger} A.,  2012, \mn@doi [\apj]
  {10.1088/0004-637X/752/1/32}, \href
  {https://ui.adsabs.harvard.edu/abs/2012ApJ...752...32W} {752, 32}

\bibitem[\protect\citeauthoryear{{Yost}, {Harrison}, {Sari}  \& {Frail}}{{Yost}
  et~al.}{2003}]{2003ApJ...597..459Y}
{Yost} S.~A.,  {Harrison} F.~A.,  {Sari} R.,   {Frail} D.~A.,  2003, \mn@doi
  [\apj] {10.1086/378288}, \href
  {https://ui.adsabs.harvard.edu/abs/2003ApJ...597..459Y} {597, 459}

\bibitem[\protect\citeauthoryear{{Zhang} \& {M{\'e}sz{\'a}ros}}{{Zhang} \&
  {M{\'e}sz{\'a}ros}}{2001}]{zhang2001}
{Zhang} B.,  {M{\'e}sz{\'a}ros} P.,  2001, \mn@doi [\apjl] {10.1086/320255},
  \href {http://adsabs.harvard.edu/abs/2001ApJ...552L..35Z} {552, L35}

\bibitem[\protect\citeauthoryear{{Zhang}, {Fan}, {Dyks}, {Kobayashi},
  {M{\'e}sz{\'a}ros}, {Burrows}, {Nousek}  \& {Gehrels}}{{Zhang}
  et~al.}{2006}]{2006ApJ...642..354Z}
{Zhang} B.,  {Fan} Y.~Z.,  {Dyks} J.,  {Kobayashi} S.,  {M{\'e}sz{\'a}ros} P.,
  {Burrows} D.~N.,  {Nousek} J.~A.,   {Gehrels} N.,  2006, \mn@doi [\apj]
  {10.1086/500723}, \href {http://adsabs.harvard.edu/abs/2006ApJ...642..354Z}
  {642, 354}

\makeatother
\end{thebibliography}




\newpage
\appendix

\begin{table}[h]
\centering
\caption{Proposed analytical functions to mimic early magnetic field growth in magnetars.}
\label{tab:math_functions}
\begin{tabular}{|l|l|}
\hline
\textbf{Function Type}        & \textbf{Mathematical Expression} \\ \hline
Exponential & \( f_1(t) = \varepsilon + \left(1 - e^{-\frac{t}{\tau_B}}\right) \) \\ \hline
Hyperbolic  & \( f_2(t) = \varepsilon + \tanh\left(\frac{t}{\tau_B}\right) \) \\ \hline
Power Law   & \( f_3(t) = \varepsilon + \frac{\left(\frac{t}{\tau_B}\right)}{\sqrt{1+ \left(\frac{t}{\tau_B}\right)^2}} \) \\ \hline
\end{tabular}
\end{table}

\begin{table}
\centering \renewcommand{\arraystretch}{1.85}\addtolength{\tabcolsep}{1.5pt}
\caption{Spectral and temporal PL indexes of synchrotron flux ($F_\nu \propto t^{-\alpha_c}\nu^{-\beta_c}$) of the afterglow model without the reemergence of magnetic field.}
\label{Table2}
{\begin{tabular}{c c c c}
\hline \hline
Frequency range & \hspace{0.5cm} $\beta_c$ & \hspace{0.5cm} $\alpha_c$ & \hspace{0.5cm} $\alpha_c(\beta_c)$ \\ \hline \hline

$\nu < \nu_{\rm a,3}$   	                                 & \hspace{0.5cm} $-2 $     & $-1$ & \hspace{0.5cm} $\frac{\beta_c}{2}$ \\
$ \nu_{\rm a,3} < \nu < \nu_{\rm c} $   	                & \hspace{0.5cm} $-\frac13$   & $-\frac{10+11\alpha}{3(\alpha+5)} $ & \hspace{0.5cm} $\frac{(10+11\alpha)\beta_c}{\alpha+5}$ \\	
$\nu_{\rm c} < \nu < \nu_{\rm m} $   	                & \hspace{0.5cm} $\frac{1}{2} $  & $-\frac{5+4\alpha}{2(\alpha+5)} $ & \hspace{0.5cm} $-\frac{(5+4\alpha)\beta_c}{\alpha+5}$ \\ 	
$\nu_{\rm m} < \nu $   	                                 & \hspace{0.5cm} $\frac{p}{2} $  & $\frac{15p-4(\alpha+5)}{2(\alpha+5)}$ & \hspace{0.5cm} $\frac{15\beta_c-2(\alpha+5)}{\alpha+5}$ \\ \hline

$\nu < \nu_{\rm a,1} $   	                                 & \hspace{0.5cm} $-2 $    & $-\frac{2(\alpha-1)}{\alpha+5}$ & \hspace{0.5cm} $\frac{(\alpha-1)\beta_c}{\alpha+5}$ \\
$ \nu_{\rm a,1} < \nu < \nu_{\rm m} $   	        & \hspace{0.5cm} $-\frac{1}{3}$    & $-\frac{8+3\alpha}{\alpha+5} $ & \hspace{0.5cm} $\frac{3\beta_c(8+3\alpha)}{\alpha+5}$ \\	
$ \nu_{\rm m} < \nu < \nu_{\rm c} $   	                & \hspace{0.5cm} $\frac{p-1}{2} $ & $\frac{3(5p-2\alpha-7)}{2(\alpha+5)}$ & \hspace{0.5cm} $\frac{3(5\beta_c-1-\alpha)}{\alpha+5}$ \\ 	
$\nu_{\rm c} < \nu $   	                                 & \hspace{0.5cm} $\frac{p}{2} $   & $\frac{15p-4(\alpha+5)}{2(\alpha+5)}$ & \hspace{0.5cm} $\frac{15\beta_c-2(\alpha+5)}{\alpha+5}$ \\ \hline

$\nu < \nu_{\rm m} $   	                                 & \hspace{0.5cm} $-2 $     & $-\frac{2(\alpha-1)}{\alpha+5}$ & \hspace{0.5cm} $\frac{(\alpha-1)\beta_c}{\alpha+5}$ \\
$ \nu_{\rm m} < \nu < \nu_{\rm a,2} $   	        & \hspace{0.5cm} $-\frac52$     & $-\frac{11+4\alpha}{2(\alpha+5)} $ & \hspace{0.5cm} $\frac{(11+4\alpha)\beta_c}{5(\alpha+5)}$ \\	
$ \nu_{\rm a,2} < \nu < \nu_{\rm c} $   	                & \hspace{0.5cm} $\frac{p-1}{2} $ & $\frac{3(5p-2\alpha-7)}{2(\alpha+5)}$ & \hspace{0.5cm} $\frac{3(5\beta_c-1-\alpha)}{\alpha+5}$ \\ 	
$\nu_{\rm c} < \nu $   	                                 & \hspace{0.5cm} $\frac{p}{2} $  & $\frac{15p-4(\alpha+5)}{2(\alpha+5)}$ & \hspace{0.5cm} $\frac{15\beta_c-2(\alpha+5)}{\alpha+5}$ \\ \hline

\end{tabular}}
\end{table}

\begin{table}
\centering \renewcommand{\arraystretch}{1.85}\addtolength{\tabcolsep}{1.5pt}
\caption{Spectral and temporal PL indexes of synchrotron flux ($F_\nu \propto t^{-\alpha_c}\nu^{-\beta_c}$)  of the afterglow model with the reemergence of magnetic field.}
\label{Table3}
{\begin{tabular}{c c c c}
\hline \hline
Frequency range & \hspace{0.5cm} $\beta_c$ & \hspace{0.5cm} $\alpha_c$ & \hspace{0.5cm} $\alpha_c(\beta_c)$ \\ \hline \hline

$\nu < \nu_{\rm a,3}$   	                                 & \hspace{0.5cm} $-2 $     & $-1$ & \hspace{0.5cm} $\frac{\beta_c}{2}$ \\
$ \nu_{\rm a,3} < \nu < \nu_{\rm c} $   	                & \hspace{0.5cm} $-\frac13$   & $\frac{5(11-5n)+11\alpha(1-n)}{3(n-1)(\alpha+5)}$ & \hspace{0.5cm} $-\frac{[5(11-5n)+11\alpha(1-n)]\beta_c}{(n-1)(\alpha+5)}$ \\	
$\nu_{\rm c} < \nu < \nu_{\rm m} $   	                & \hspace{0.5cm} $\frac{1}{2} $  & $\frac{5(2-n)+2\alpha(1-n)}{(n-1)(\alpha+5)}$ & \hspace{0.5cm} $\frac{2[5(2-n)+2\alpha(1-n)]\beta_c}{(n-1)(\alpha+5)}$ \\ 	
$\nu_{\rm m} < \nu $   	                                 & \hspace{0.5cm} $\frac{p}{2} $  & $\frac{2(\alpha+5)(1-n)+5np}{(n-1)(\alpha+5)}$ & \hspace{0.5cm} $\frac{2[(\alpha+5)(1-n)+5n\beta_c]}{(n-1)(\alpha+5)}$ \\ \hline

$\nu < \nu_{\rm a,1} $   	                                 & \hspace{0.5cm} $-2 $    & $\frac{2[5-n+\alpha(1-n)]}{(n-1)(\alpha+5)}$ & \hspace{0.5cm} $-\frac{[5-n+\alpha(1-n)]\beta_c}{(n-1)(\alpha+5)}$ \\
$ \nu_{\rm a,1} < \nu < \nu_{\rm m} $   	        & \hspace{0.5cm} $-\frac{1}{3}$    & $\frac{9(\alpha+5)-n(31+9\alpha)}{3(n-1)(\alpha+5)}$ & \hspace{0.5cm} $-\frac{[9(\alpha+5)-n(31+9\alpha)]\beta_c}{(n-1)(\alpha+5)}$ \\	
$ \nu_{\rm m} < \nu < \nu_{\rm c} $   	                & \hspace{0.5cm} $\frac{p-1}{2} $ & $\frac{3(\alpha+5)+5np-3n(\alpha+4)}{(n-1)(\alpha+5)}$ & \hspace{0.5cm} $\frac{3(\alpha+5)+10n\beta-n(3\alpha+7)}{(n-1)(\alpha+5)}$ \\ 	
$\nu_{\rm c} < \nu $   	                                 & \hspace{0.5cm} $\frac{p}{2} $   & $\frac{2(\alpha+5)(1-n)+5np}{(n-1)(\alpha+5)}$ & \hspace{0.5cm} $\frac{2[(\alpha+5)(1-n)+5n\beta_c]}{(n-1)(\alpha+5)}$ \\ \hline

$\nu < \nu_{\rm m} $   	                                 & \hspace{0.5cm} $-2 $     & $\frac{2[5-n+\alpha(1-n)]}{(n-1)(\alpha+5)}$ & \hspace{0.5cm} $-\frac{[5-n+\alpha(1-n)]\beta_c}{(n-1)(\alpha+5)}$ \\
$ \nu_{\rm m} < \nu < \nu_{\rm a,2} $   	        & \hspace{0.5cm} $-\frac52$     & $\frac{2(\alpha+5)-n(7+2\alpha)}{(n-1)(\alpha+5)}$ & \hspace{0.5cm} $-\frac{2[2(\alpha+5)-n(7+2\alpha)]\beta_c}{5(n-1)(\alpha+5)}$ \\	
$ \nu_{\rm a,2} < \nu < \nu_{\rm c} $   	                & \hspace{0.5cm} $\frac{p-1}{2} $ & $\frac{3(\alpha+5)+5np-3n(\alpha+4)}{(n-1)(\alpha+5)}$ & \hspace{0.5cm} $\frac{3(\alpha+5)+10n\beta_c-n(3\alpha+7)}{(n-1)(\alpha+5)}$ \\ 	
$\nu_{\rm c} < \nu $   	                                 & \hspace{0.5cm} $\frac{p}{2} $  & $\frac{2(\alpha+5)(1-n)+5np}{(n-1)(\alpha+5)}$ & \hspace{0.5cm} $\frac{2[(\alpha+5)(1-n)+5n\beta_c]}{(n-1)(\alpha+5)}$ \\ \hline

\end{tabular}}
\end{table}

\clearpage

\begin{table}
\centering
\caption{Parameter values used (Figures \ref{fig:AfterglowVar} - \ref{fig:MagVarRela}) and obtained (Figure \ref{fig:GRB170817A}) for the synchrotron light curves at X-ray, optical and radio bands generated  by the deceleration of the non-relativistic material in the external environment.}
\label{tab:par_values}
\begin{tabular}{|l|l|}
\hline \hline
\textbf{Parameter}        & \hspace{1.6cm}\textbf{Values} \\
  &   General \hspace{1.5cm} GRB 170817A \\ 
  & (Figs. \ref{fig:AfterglowVar} - \ref{fig:MagVarRela})\hspace{1.2cm} (Fig. \ref{fig:GRB170817A})\\ \hline \hline
Emergence of B-field model&  \\ \hline
$\tau_B$\, (yr) &  $3$\hspace{2.3cm} $5.5\pm 0.2$\\
$\epsilon$\, &  $10^{-4}$ \hspace{1.8cm} $(1.1\pm 0.3)\times 10^{-4}$\\
$\tau_0$\, (yr) &  $2.5$ \hspace{2cm} $10.5\pm0.1$\\
$\dot{E}_0$\, (erg/s) &  $4.5\times 10^{43}$ \hspace{1.1cm} $(1.5\pm 0.2)\times 10^{41}$ \\\hline

Afterglow scenario&  \\ \hline
$E$\, (erg) &  $10^{49}$ \hspace{1.75cm} $(6.2\pm0.5)\times 10^{49}$\\
$n_s$\, (cm$^{-3}$) &  $1$ \hspace{2.1cm} $(2.8\pm 0.4)\times 10^{-4}$\\
$\varepsilon_{\rm e}$\, &  $10^{-1}$ \hspace{1.7cm} $(9.3\pm 1.0)\times 10^{-2}$\\
$\varepsilon_{\rm B}$\, &  $10^{-2}$ \hspace{1.7cm} $(6.9\pm 0.5)\times 10^{-4}$\\
$p$\, &  $2.5$ \hspace{2cm}$2.2\pm 0.1$\\
$\alpha$\, &  $2.2$ \hspace{2cm}$3.0\pm 0.1$\\
$\theta_j$ (deg)\, &  $4.6$ \hspace{2cm}$7.5\pm 0.3$\\
$\Delta\theta$ (deg)\, &  $15.0$ \hspace{1.85cm}$18.8\pm0.3$\\\hline
\end{tabular}
\end{table}

\newpage

\bsp	
\label{lastpage}
\end{document}